\documentclass{./sig-alternate}

\usepackage{times}
\usepackage{subfigure}
\usepackage{amsmath}
\usepackage{amssymb}
\usepackage{color}
\usepackage{array}
\usepackage{pifont}
\usepackage[ruled,linesnumbered,vlined]{algorithm2e}
\usepackage[normalem]{ulem}

\usepackage{verbatim}
\usepackage{graphics}
\usepackage[hyphens]{url}
\usepackage[breaklinks=true]{hyperref}
\usepackage[hyphenbreaks]{breakurl}
\usepackage[table]{xcolor}
\usepackage[T1]{fontenc}
\usepackage{mathtools}
\usepackage{mhchem}
\usepackage[slovak,english]{babel}
\usepackage{graphicx}
\usepackage{balance} 
\usepackage{soul}
\usepackage{enumitem}
\usepackage{amsmath}

\usepackage{subfigure}
\usepackage{epsfig}
\usepackage{epstopdf}
\usepackage{pifont}
\usepackage{verbatim}

\usepackage[hyphens]{url}
\usepackage[breaklinks=true]{hyperref}
\usepackage[hyphenbreaks]{breakurl}
\title{EAGr: Supporting Continuous \uline{E}go-centric \uline{A}ggregate Queries over Large Dynamic \uline{Gr}aphs}

\author{Jayanta Mondal\qquad  Amol Deshpande\\\\
\textsf{\normalsize{Dept. of Computer Science, University of Maryland, College Park, MD 20742}}\\  
\vspace{4pt}
\textsf{\normalsize{\{jayanta, amol\}@cs.umd.edu}}}

\newcommand{\calG}{\mathcal{G}}
\newcommand{\calN}{\mathcal{N}}
\newcommand{\calF}{\mathcal{F}}

\newcommand{\calI}{\mathcal{I}}

\newcommand{\amolnote}[1]{\noindent{\textcolor{red}{\bf Amol note: #1}}}
\newcommand{\jayantanote}[1]{\noindent{\textcolor{blue}{Jayanta: #1}}}

\newcommand{\red}[1]{\noindent{\textcolor{red}{#1}}}

\newcommand{\topicu}[1]{\vspace{8pt} \noindent \underline{\bfseries #1:}}
\definecolor{mygreen}{rgb}{.1,.5,.1}
\definecolor{mygray}{rgb}{.5,0.5,.5}
\definecolor{mygray2}{rgb}{.8,0.8,.8}
\newcommand{\jayanta}[1]{\noindent{\textcolor{mygray}{Jayanta: #1}}}

\newcommand{\suma}{{\sc Sum}}
\newcommand{\mina}{{\sc Min}}
\newcommand{\maxa}{{\sc Max}}
\newcommand{\counta}{{\sc Count}}
\newcommand{\topka}{{\sc Top-K}}
\newcommand{\uniquea}{{\sc Unique}}

\newcommand{\vnm}{{\sc Vnm}}
\newcommand{\vnma}{{\sc Vnm}$_A$}
\newcommand{\vnmn}{{\sc Vnm}$_N$}
\newcommand{\vnmd}{{\sc Vnm}$_D$}
\newcommand{\iob}{{\sc Iob}}

\newtheorem{theorem}{Theorem}[section]

\newcommand{\eat}[1]{}

\newcommand{\squishlist}{
 \begin{list}{$\bullet$}
  { \setlength{\itemsep}{2pt}
    \setlength{\parsep}{2pt}
     \setlength{\topsep}{2pt}
     \setlength{\partopsep}{2pt}
     \setlength{\leftmargin}{1.5em}
     \setlength{\labelwidth}{1em}
     \setlength{\labelsep}{0.5em} } }

\newcommand{\squishlisttwo}{
 \begin{list}{$\bullet$}
  { \setlength{\itemsep}{2pt}
     \setlength{\parsep}{2pt}
    \setlength{\topsep}{2pt}
    \setlength{\partopsep}{2pt}
    \setlength{\leftmargin}{2em}
    \setlength{\labelwidth}{1.5em}
    \setlength{\labelsep}{0.5em} } }

\newcommand{\squishend}{
  \end{list}  }

\begin{document}
\raggedbottom


\maketitle


\begin{abstract}
In this paper, we present EAGr, a system for supporting large numbers of continuous
neighborhood-based (``\uline{e}go-centric'') \uline{a}ggregate queries over large, highly dynamic, and rapidly evolving 
\uline{gr}aphs. 
Examples of such queries include computation of {\em personalized, tailored trends}
in social networks,
{\em anomaly or event detection} in communication or
financial
transaction networks, {\em local search} and {\em alerts} in spatio-temporal networks, to name a few.
Key challenges in supporting such continuous queries include very high update rates 
typically seen in these situations, large numbers of queries that need to be executed simultaneously, 
and stringent low latency requirements. In this paper, we propose a flexible, general, and extensible
in-memory framework for executing different types of ego-centric aggregate queries over large dynamic 
graphs with low 
latencies. Our framework is built around the notion of an {\em aggregation overlay graph}, a 
pre-compiled data structure that encodes the computations to be performed when an update or
a query is received. The overlay graph enables {\em sharing of partial aggregates} across different
ego-centric queries (corresponding to different nodes in the graph), and also allows {\em partial
pre-computation} of the aggregates to minimize the query latencies. 
We present several highly scalable techniques for constructing an overlay graph given 
an aggregation function, and also design incremental algorithms for
handling changes to the structure of the underlying graph itself, that may result in significant
changes to the neighborhoods on which queries are posed. We also present an optimal, polynomial-time
algorithm for making the pre-computation decisions given an overlay graph, and evaluate an approach
to incrementally adapt those decisions as the workload changes.
Although our approach is naturally parallelizable, we focus on a single-machine deployment in this paper and show that our
techniques can easily handle
graphs of size up to 320 million nodes and edges, and achieve update and query throughputs of over 500,000/s using a single, powerful machine.

\end{abstract}





\newpage
\section{Introduction}


Graph-structured data arises naturally in a variety of application domains, including social
networks, communication networks, phone call networks, email networks, financial transaction
networks, to name a few. There is an increasing need to support graph structure-aware queries
and analysis tasks over such graphs, leading to much work in this area over the last few years.
In many of these domains, the datasets are not only large in terms of the sheer number of 
nodes and edges in the graph, but they also produce a large amount of data at a very high rate, 
      generating a {\em data stream} that must be ingested and queried in real time.
The graph data can be seen as comprising of two major components: (a) a graph ({\em network}) component that captures the underlying interconnection structure among the
nodes in the graph, and (b) {\em content} data associated with the nodes and the edges.
The graph data stream contains updates to both these components. The structure of the graph may itself change
rapidly in many cases, especially when things like webpages, user tags (e.g., Twitter {\em hashtags}), financial trades,
etc., are treated as nodes of the graph. However, most of the data stream consists of updates to
the content data associated with the nodes and edges, e.g., status updates or photos uploaded by
social network users, phone calls or messages among users, transactions in a financial network, etc. 
Real-time, continuous query processing over such dynamic graph-structured data has become a critical 
need in the recent years. 

In this paper, we focus on a prevalent class of queries over dynamic graphs, called {\em
neighborhood-based or ego-centric aggregate queries}. In an ego-centric aggregate query, the
querier (called {\em user} henceforth) corresponds to a node in the graph, and is interested 
in an aggregate over the current state or the recent history of a local neighborhood of the node in 
the graph; such local neighborhoods are often called {\em ego networks} of the nodes~\cite{egonetLeskovec, egonetEverett}. 
An example of such a query is {\em
ego-centric trend analysis in social networks} where the goal is to find, for each user, the trends 
(e.g., popular topics of discussion, news items) in his or her local neighborhood~\cite{abel2011analyzing, guy2010social}. The neighborhood
here could be 1-hop neighborhood, or could extend beyond that. Similarly, in a phone-call network or
an analogous communication network, we may be interested in identifying interesting events or anomalies (e.g., higher than
normal communication activity among a group of nodes);
that often boils down to continuously computing ego-centric aggregates over recent activity 
in a large number of local neighborhoods simultaneously (with an anomaly defined by a predicate on the aggregate)~\cite{oddball,bgplens}.
In spatio-temporal social networks, users are often interested in events happening in their 
social networks, but also physically close to them.

We make a distinction between between {\em continuous} queries and what we call {\em
quasi-continuous} queries (somewhat surprisingly, we have not seen this distinction made in prior
work). In the latter case, the query result only needs to produced or updated when the user requests
it (we call such user requests {\em reads}); whereas in the former case, the query result must be 
kept up-to-date whenever the inputs change.
The first query above (trend analysis) is an example of a quasi-continuous query since there is no 
need to produce the result unless the user asks for it (for reducing latency, full or partial pre-computation 
may be performed). However, anomaly detection queries must be executed continuously as
new updates arrive.

The high update rates typically seen in these application domains make it a challenge to
execute a large number of such queries with sufficiently low latencies. A naive {\em
on-demand} approach, where the neighborhood is traversed in response to a read, is unlikely 
to scale to the large graph sizes, and further, would have unacceptably high query latencies. On the other hand,
a {\em pre-computation-based approach}, where the required query answers are always pre-computed and 
kept up-to-date will likely lead to much wasted computation effort for most queries. Furthermore, both 
these approaches ignore many potential optimization opportunities, in particular, the possibility of
{\em sharing} the aggregate computation across different queries (corresponding to different ego
networks).

In this paper, we propose an approach that maintains a special directed graph 
(called an {\em aggregation overlay graph} or simply an {\em overlay}) 
that is constructed given an ego-centric aggregate query and a subset of nodes in the data graph 
for which it needs to be evaluated continuously (or quasi-continuously).
The overlay graph exposes sharing opportunities by explicitly utilizing {\em partial aggregation} nodes, whose outputs can be shared
across queries.
The nodes in the overlay are labeled with {\em dataflow decisions} that encode whether data should be {\em pushed} to
that node in response to an update, or it should be {\em pulled} when a query result needs to be computed.
During execution, the overlay simply reacts to the events (i.e., reads and writes) based on the encoded decisions, and is thus
able to avoid unnecessary computation, leading to very high throughputs across a spectrum of workloads.
Constructing the optimal overlay graph is NP-Hard for 
arbitrary graph topologies. Further, given the large network sizes that are typically
seen in practice, it is infeasible to use some of the natural heuristics for solving this problem.
We present  a series of highly efficient overlay construction algorithms and show how they can
be scaled to very large graphs.
Surprisingly, the problem of making the dataflow decisions for a given overlay is solvable in polynomial time, 
and we present a {\em max-flow}-based algorithm for that purpose.
Our framework can support different
neighborhood functions (i.e., 1-hop, 2-hop neighborhoods), and also allows filtering neighborhoods (i.e., only aggregating
over subsets of neighborhoods). The framework also supports a variety of aggregation functions (e.g., {\em sum, count, min, max,
top-k}, etc.), exposes an aggregation API for specifying and executing arbitrary user-defined aggregates. %
We conduct a comprehensive experimental evaluation over a collection of 
real-world networks, our results show that
overlay-based execution of aggregation queries saves redundant computation and significantly boosts the 
end-to-end throughput of the system.


\eat{
\amolnote{Need to discuss technical details. Problems/NP-Hardness/Basic ideas. Done:yet to discuss the push/pull problem}

\amolnote{move text from this paragraph up or down. Now eaten.}\\
}
\eat{
\topicu{Our goal} In this work we aim to build a system that can support neighborhood-based aggregations on large, dynamically changing graphs in a scalable way. Keeping with the hardware trends[x], to support low latency operations, our system is designed to be fully in-memory, uses disk mainly for checkpointing. Some of the key challenges in building such system are: (1) performing neighborhood-based aggregation is computation-heavy, (2) presence of cycles might require graph duplicate detection, (3) graph structure changes dynamically. A naive approach like gossip based scheme does not work well because it requires too many graph traversals and expensive duplicate detection at run-time. These type of techniques not only increase query latencies but also increase the total amount of data movement within memory, thus limiting to the scalability of the system.
}

\vspace{5pt}
\noindent
\textbf{Outline:} We begin with a brief overview of the problem by discussing the data and the query
model (Section~\ref{sec:overview}).  Then we present the
details of our proposed aggregation framework, including the API for supporting user-defined
aggregates (Section~\ref{sec:framework}). Next we analyze 
the optimization problem of constructing an {\em overlay
graph} (Section~\ref{sec:cdfg}), and propose several efficient heuristics that can scale to very
large graphs. Following that, we discuss how we to make the dataflow (push/pull) decisions in order to minimize data movement 
in the overlay graph (Section~\ref{sec:pushpull}). Then we describe our experimental setup and present a comprehensive experimental evaluation 
(Section~\ref{sec:evaluations}), and discuss
some of the most related work (Section~\ref{sec:rltdwork}). 
\eat{Finally we conclude (Section~\ref{sec:conclusions}) and cite the references.}


\section{Overview}
\label{sec:overview}
We start with describing the underlying data and query model, followed by an overview of
our proposed aggregation framework.

\begin{table} [t]
\begin{tabular}{|p{.6in}|p{2.4in}|}
\hline
 \multicolumn{1}{|c|}{ \parbox{0.4in}{\vspace{1pt}Notation\vspace{1pt}}} &    \multicolumn{1}{c|}{\parbox{0.4in}{\vspace{1pt}Description\vspace{1pt}}}\\
  \hline
  \parbox{2in}{\vspace{2pt}$\mathcal{G}(V,E)$ }& Underlying data graph\\
  
  $\mathcal{N}()$ & Neighborhood selection function\\
 
 $\calF()$ & Aggregate function\\  
{\em write} on $v$ & An update to node $v$'s content \\
{\em read} on $v$ & A read to query result at $v$, i.e., $\calF(\mathcal{N}(v))$\\
 $A_{\calG}(V',E')$ & Bipartite directed writer/reader graph: for each node $v \in \mathcal{G}(V, E)$, it contains
 two nodes $v_w$ (writer) and $v_r$ (reader), with edges going from writers to readers \\  
 
 $O_\mathcal{G}(V'',E'')$ & Overlay Graph\\
 
 $\calI(ovl)$ & Set of writers aggregated by overlay node $ovl$ \\ 
  
 $w(v)$ & write frequency of node $v$\\
 $r(v)$ & read (query) frequency of node $v$\\
 $f_h(v)$ & push frequency of node $v$ in an overlay\\
 $f_l(v)$ & pull frequency of node $v$ in an overlay\\

 \hline
\end{tabular}
\caption{Notation}
\label{table1}
\end{table}

\begin{figure*}
\centering
\includegraphics[width=6.9in]{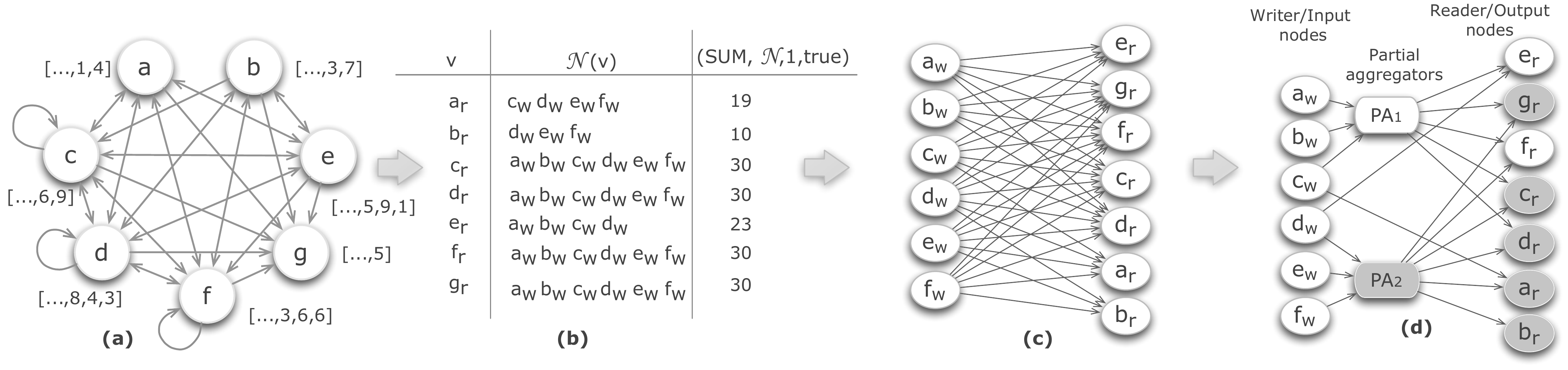}
 \caption{(a) An example data graph, (b) $\mathcal{N}(v)$ and {\sc Sum} aggregates for each $v$,
     (c) Bipartite representation of the graph, i.e, $A_{\calG}$ (note, $g$ does not form input to any reader), (d) An
         overlay graph (shaded nodes indicate {\em pull} decisions, unshaded ones indicate {\em push}).}
\label{fig:longexample}
\end{figure*}

\subsection{Data and Query Model}
\label{sec:model}
\topicu{Data Model}
Let $\mathcal{G}(V, E)$ denote the underlying
connection graph, with $V$ and $E$ denoting the sets of nodes and edges 
respectively. In general, $\mathcal{G}$ is a heterogeneous, multi-relational graph that may
contain many different types of nodes and may contain both directed and undirected edges. 
For example, for a social network, we may have nodes representing the users of the network
as well as nodes representing communities, groups, user tags, webpages, and so on.
Similarly, $E$ may include not only symmetric {\em friendship} (or analogous) edges but also asymmetric {\em
follows} edges, {\em membership} edges, and other types of semi-permanent edges that are usually in
existence from the time they are formed till the time they are deleted (or till the current
time). The {\em content} associated with the nodes and edges is captured through a set of {\em
attribute-value} pairs. 


We capture the structure updates (i.e., node or edge additions or deletions) as a time-stamped data stream
$S_{\mathcal{G}}$ (called {\em structure data stream}). For simplicity, we assume that all the content is associated with nodes, and for
a node $v$, we capture the content updates associated with it as a time-stamped data stream, $S_v$
(called {\em content data streams}). We further assume that all the content streams are homogeneous, i.e., all 
updates are of the same type or refer to the same attribute. It is straightforward 
to relax both these assumptions. A content update on node $v$ is also called a {\em write on $v$}.

Unlike most prior work in data streams or publish-subscriber networks where the producers of data
(i.e., writers) and the consumers of data (i.e., readers) are distinct from each other, in our case,
a node acts both as a writer and a reader.
Hence, for clarity of description, when referring to a node $v$ in the rest of the paper, we often denote
its role in the context using a subscript -- $v_w$ (similarly, $v_r$) denoting the node as a writer (reader).


\topicu{Query Model} An ego-centric aggregate query is specified by four parameters: 
$\langle \mathcal{F}, w, \calN{}, pred \rangle$, where $\mathcal{F}$ denotes the aggregate function to be
computed, $w$ denotes a sliding window over the content data streams, $\calN{}$ denotes the
neighborhood selection function (i.e., $\calN{(v)}$ forms the input list to be aggregated for each $v$) , and $pred$ selects a subset of
$V$ for which the aggregate must be computed (i.e., $\mathcal{F}$ would be computed for all nodes for which $pred(v)$ is {\em true}).
Following the data streams literature, $w$ may be a time-based sliding window or a tuple-based sliding
window; in the former case, we are given a time interval $T$, and the updates that arrive within the
last $T$ time are of interest, whereas in the latter case, we are given a number $c$, and the last
$c$ updates are of interest. The query may be specified to be a
{\em continuous} query or a {\em quasi-continuous} query. For a continuous query, the query results
must be kept up-to-date as new updates arrive, whereas for a quasi-continuous query, the query
result for a node $v$ is only needed when a user requests it (we call this a {\em read on} $v$); in the latter case,
pre-computation may be done to reduce not only user latencies but also total computational effort.

Since our approach is based on pre-computation and maintenance of partial aggregates, we 
assume that the aggregate function (and $\calN{}$) are pre-specified. In some cases, it is possible to
share the intermediate data structures and partial aggregates for simultaneous evaluation of
different aggregates; we do not consider that option further in this paper. 

Our system supports a set of built-in aggregate functions like {\em sum, max, min, top-k, etc.},
and we allow the user to define arbitrary aggregation functions (Section \ref{sec:api}). Our 
system treats $\mathcal{F}$ as a blackbox, but the user may optionally specify whether the 
aggregation function is {\em duplicate-insensitive} or supports efficient {\em subtraction} (Section~\ref{problem}), and that information will be used to further
optimize the computation.

\topicu{Example} Figure \ref{fig:longexample} shows an example instance of this problem. 
Figure \ref{fig:longexample}(a) depicts the data graph. $\mathcal{N}(x)$ is defined to be $\{y  | y \rightarrow x \}$
(note that, all edges are not bidirectional).
The numbers in the square brackets denote individual content streams. 
For example, there have been two {\em recent writes} on node $a$ with values $1$ and $4$.
The query is $\langle $\suma$, c=1, \calN{}, v\in V \rangle$, which states that for each node $v \in V$, the most recent
values written by nodes in
  $\mathcal{N}(v)$ need to be aggregated using \suma{}.
Figure~\ref{fig:longexample}(b) enumerates $\mathcal{N}(v)$ for each $v$. 
The last column of Figure~\ref{fig:longexample}(b) shows the results of the {\em read} queries on
each node. For example, here $\calN{(a)}$ evaluates to $\{c, d, e, f\}$, and a {\em read} query on $a$ returns: $(9)+(3)+(1)+(6)=19$. 
Figure~\ref{fig:longexample}(c) represents the corresponding directed bipartite graph $A_\calG$ where nodes are duplicated
and divided based on their roles;
a node might or might not play both the roles. 

\topicu{Scope of the Approach} Here, we briefly summarize the key assumptions that we make and the limitations of our approach.
Our compilation-based approach requires upfront knowledge of the query to be executed, including the specific aggregate
function, the neighborhood to aggregate over, and the sliding window parameters (the last of which only impacts dataflow
decisions). Further, given the high overlay construction cost, the query needs to be evaluated continuously for a period
of time to justify the cost. Thus our approach would not be suitable for ad hoc ego-centric aggregate queries over graphs.
We also note that, although our framework can handle arbitrary aggregation functions, the benefits of our approach, especially of
sharing partial aggregates, are higher for {\em distributive} and {\em algebraic} aggregates than for {\em holistic}
aggregates like {\em median}, {\em mode}, or {\em quantile} (however, approximate versions of holistic aggregate can still benefit a from our optimizations). Our approach to making dataflow decisions based on expected read/write frequencies also requires the ability to estimate or predict those frequencies. As with most workload-aware approaches, our approach will likely not work well in face of highly unpredictable and volatile workloads. Finally, we also assume that the data graph itself changes relatively slowly; although we have developed incremental techniques to modify the overlay in such cases, our approach is not intended for scenarios where the structure of the data graph changes rapidly.

\subsection{Proposed Aggregation Framework} 
\label{sec:framework}

In this section, we describe our proposed framework to support different types of ego-centric aggregate 
queries. We begin with explaining the notion of an {\em aggregation overlay graph} and key rationale behind it. 
We then discuss the execution model and some of the key implementation issues. 

\subsubsection{Aggregation Overlay Graph}
\label{sec:model-overlay}

Aggregation overlay graph is a pre-compiled data structure built for a given ego-centric aggregate
query, that enables sharing of partial aggregates, selective pre-computation, partial pre-computation, and low-overhead query
execution. Given a data graph $\mathcal{G}(V, E)$ and a query $\langle \mathcal{F}, w, \calN{},$ $ pred \rangle$, 
we denote an aggregation overlay graph for them by $O_\mathcal{G}(V'',E'')$. 


There are three types of nodes in an overlay graph: (1) the {\em writer} nodes, denoted by subscript
${\_}_w$, one for each node in
the underlying graph that is {\em generating} data, (2) the {\em reader} nodes, denoted by subscript
${\_}_r$, 
one for each node in $V$ that satisfies $pred$, and (3) the {\em partial aggregation} nodes (also
called {\em intermediate} nodes). 
We use the term {\em aggregation} node to refer to either a reader node or a partial aggregation
node, since both of those may perform aggregation.
In Figure \ref{fig:longexample}(d), $PA_1$ and $PA_2$ are two partial aggregation nodes that 
are introduced after analyzing the structure of the network and the query. $PA_1$ corresponds to
a partial aggregator that aggregates the inputs $a_w, b_w$, $c_w$, and serves  $e_r$, $g_r$,  $f_r$, $c_r$, $d_r$.

For correctness, there can only be one (directed) path from a writer to a reader in an overlay graph (to
avoid duplicate contributions from that writer to the aggregate computed for that reader). However, there are 
two exceptions to this. First, this is not an issue with the so-called {\em duplicate-insensitive} aggregates 
like \maxa{}, \mina, \uniquea{}. We exploit this by constructing overlays that allow such multiple 
paths paths for those aggregates, if it leads to smaller overlays (in most cases, we observed that to be the case).


Second, we allow an overlay to contain what we call {\em negative} edges to ``subtract'' such duplicate 
contributions.
A negative edge from a node $u$ to
an aggregation node $v$ indicates that the input from $u$ should be ``subtracted'' (i.e., its
contribution removed) from the aggregate result computed by $v$. Such edges should only be used 
when the ``subtraction'' operation is efficiently computable. Although negative
edges may appear to lead to wasted work, in practice, adding negative edges (where permissible) 
can actually lead to significant improvements in the total throughput. We discuss this issue 
further in Section~\ref{problem}.

The overlay graph also encodes pre-computation decisions (also called {\em dataflow decisions}). Each node 
in the overlay graph is annotated either {\em pull} or {\em push}. If a node is 
annotated {\em push}, the partial aggregate that it computes is always kept up-to-date as new updates
arrive. The writer nodes are always annotated {\em push}. For an aggregation node to be annotated {\em
push}, all its input nodes must also be annotated {\em push}. Analogously, if a node is annotated
{\em pull}, all the nodes downstream of it must also be annotated {\em pull}.
In Figure
\ref{fig:longexample}(d), the push and pull decisions are shown with unshaded and shaded nodes respectively. 
This overlay graph fully pre-computes the query results for nodes $e_r$ and $f_r$ (thus leading to low latencies for
those queries); on the other hand, a read on node $g_r$ will incur a high latency since the computation will be
done fully on demand.

Note that, we require that the decisions be made for each node in the overlay graph, rather than
for each edge. Thus, all the inputs to an aggregation node are either pushed to it, or all the inputs
are pulled by it. This simplifies the bookkeeping significantly, without limiting the choices of partial
pre-computation. If we desire to pre-compute a partial aggregate over a subset of a node's inputs,
a separate partial aggregation node can be created instead. We discuss more details about this in Section ~\ref{sec:pushpull}.

Finally, we note that the aggregation overlay graph can be seen as a pre-compiled query plan where
no unnecessary computation or reasoning is performed when an update arrives or a read query is 
posed. This enables us to handle much higher data rates than would be possible otherwise. We discuss the
resulting execution model and related architectural decisions in the following sections. 

\subsubsection{Execution Model}
\label{execmodel}
We begin with describing how new updates and queries are processed using the overlay graph, and briefly discuss some of
the implementation issues surrounding multi-threaded execution.

\topicu{Execution Flow}
We describe the basic execution flow in terms of the {\em partial aggregate objects (PAOs)} that are
maintained at various nodes in the overlay graph. A PAO corresponds to a partial aggregate that has
been computed after aggregating over a subset of the inputs.  The PAO corresponding to a node
labeled {\em push} is always kept up-to-date as new updates arrive in any of the streams it is
defined over, or if the sliding windows shift and values drop out of the window.
Specifically, the updates originate at the writer nodes, 
and propagate through the overlay graph as far as indicated by the dataflow decisions on the nodes.
The nodes labeled {\em push} maintain partial state and perform incremental computation to keep
their corresponding PAOs up-to-date. On the other hand, no partial state is maintained at the nodes 
labeled {\em pull}. 
When an overlay node $u$ makes a {\em read} request from another node $v$ upstream of it, if $v$ is
labeled {\em push}, the partial aggregate is returned immediately without delay. On the other hand,
if $v$ is labeled {\em pull}, it issues {\em read} requests on all its upstream overlay nodes,
merges all the PAOs it receives, and returns the result PAO to the requesting node.

%

\topicu{Single-threaded Execution} A naive implementation of the above execution model is using a
single thread, that processes the {\em writes} and {\em reads} in the order in which they are
received, finishing each one fully (i.e., pushing the writes as far as required, and computing
the results for {\em reads}) before handling the next one. The main advantage of this model is that
the partial state maintained at the overlay nodes and the query results generated, are both
well-defined and consistent (ignoring the temporary inconsistencies while an update is being pushed
through the overlay). However, this approach cannot exploit the parallelism in the system, is likely
to suffer from potential cache misses due to the random access pattern, and is unlikely to scale to the 
high update and query rates seen in practice.

\topicu{Multi-threaded Execution} On the other hand, a multi-threaded version can result in better
throughputs and latencies, but requires careful implementation to guarantee correctness.
First, the computations on the overlay graph must be made thread-safe to avoid potential state
corruption due to race conditions. We can do this either by using thread-safe data structures to
store the PAOs or through explicit synchronization. We use the latter approach in our implementation
of the aggregates; however, user-defined aggregates may choose either of the two options.
A more subtle issue is that of consistency. For example, consider a read on node $a_r$ in Figure
\ref{fig:longexample}(d). It is possible that the result generated for the query contains a more recent update on
node $f_w$, but does not see a relatively older update on node $c_w$ (as $f_w$ is read
later than $c_w$). We ignore the potential
for such inconsistencies in this work and plan to address this in future.

We use two thread pools, one for servicing the read requests and one for servicing the write
requests. The relative sizes of the two thread pools can be set based on the expected number of
reads vs writes; assigning more threads to processing reads may reduce latency, but increases the possibility of stale 
results.



Further, there are two ways to process a read or a write using multiple threads. The first option is what
we call the {\em uni-thread} model -- here a thread that picks up a request (read or write) executes
it fully before processing a new request. Alternatively, in the {\em queueing} model, the tasks are
subdivided into micro-tasks at the granularity of the overlay nodes. Each micro-task is responsible
for a single partial aggregate update operation at an overlay node (for writes) or a single partial
aggregate computation at an overlay node (for reads). 
The queueing model is likely to be more scalable and result
in better throughputs, but the latencies for reads are substantially higher. We follow a
hybrid approach -- for writes, we use the queueing model, whereas for reads, we use the uni-thread
model.

\subsubsection{User-defined Aggregate API} 
\label{sec:api}
One of the key features of our system is the ability for the users to define their own aggregate
functions. We build upon the standard API for user-defined aggregates for this
purpose~\cite{hellerstein2012madlib,yu2009sosp,tag}, briefly describe
it here for completeness. The user must implement the following functions. 
\squishlist
        \item {\sc initialize({\em PAO})}: Initialize the requisite data structures to maintain the
        partial aggregate state (i.e., PAOs).
        \item {\sc update({\em PAO, PAO\_old, PAO\_new})}: This is the key function that updates the
        partial aggregate at an overlay node (PAO) given that one of its inputs was updated from {\em
        PAO\_old} to {\em PAO\_new}.
		\item {\sc finalize({\em PAO})}: Compute the final answer from the PAO. 
\squishend
Note that, we require the ability to {\em merge} two PAOs in order to fully exploit the potential
for sharing through overlay graphs -- this functionality is typically optional in user-defined
aggregate APIs.

\begin{figure*}[t]
\centering
  \includegraphics[width=4.8in]{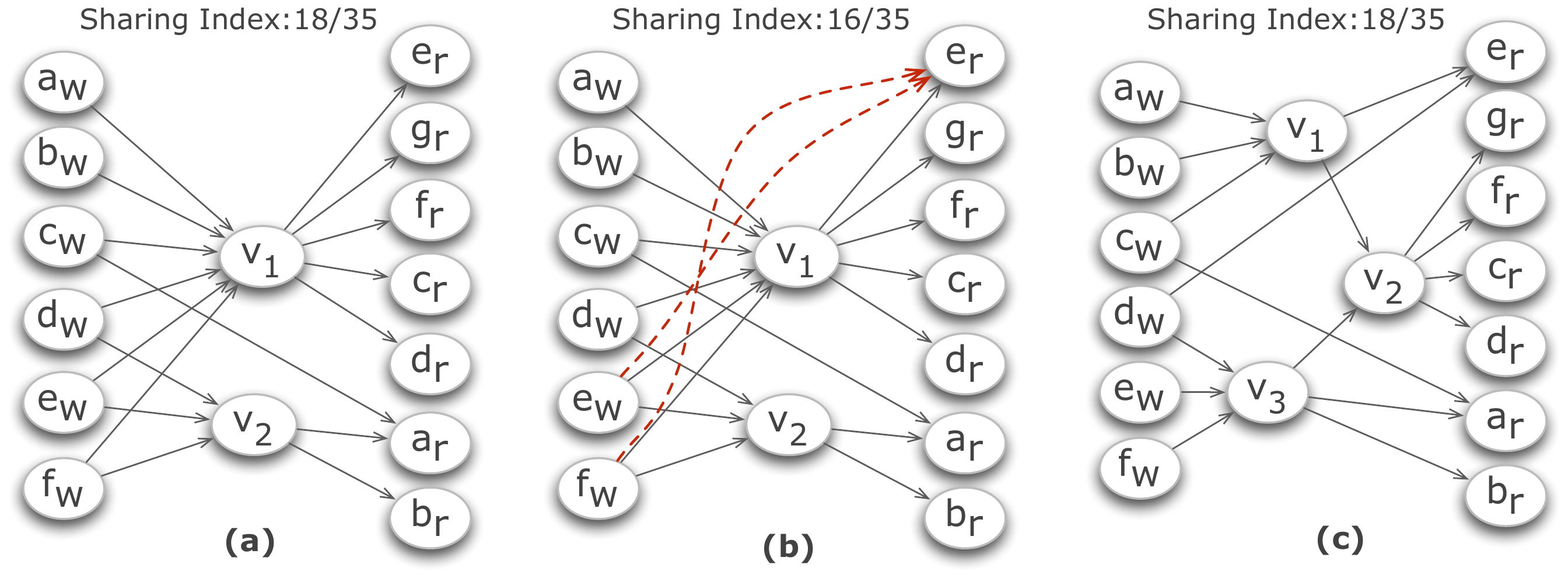}
  \vspace{-5pt}
\caption{(a) A duplicate-insensitive overlay; (b) An overlay with two negative
    edges; (c) A multi-level overlay.}
\label{fig:model-example}
\end{figure*}

\section{Constructing The Overlay}
\label{sec:cdfg}
Our overall optimization goal is to construct an overlay graph annotated with pre-computation (dataflow) decisions that
maximize the overall throughput, given a data graph and an ego-centric aggregate query. To make the 
dataflow decisions optimally, we also need information about the expected read (query) and write (update) frequencies for the nodes in the graph. However, these two 
sets of inputs 
have inherently different dynamics -- the data graph is expected to change relatively
slowly, whereas the read/write frequencies are expected to show high variability over time. Hence, we decouple the overall problem
into two phases:
%
(1) we construct a compact overlay that maximizes the sharing opportunities given a data graph and a query, and 
(2) then make the dataflow decisions for the overlay nodes (as we discuss in the next section, we allow the second phase
to make restricted local modifications to the overlay). 
The overlay construction is a computationally expensive process, and we expect
that an overlay, once constructed, will be used for a long period of time  (with incremental local changes to handle new nodes or edges).
On the other hand, we envision re-evaluating the dataflow decisions on a more frequent basis by continuously monitoring the 
read/write frequencies to identify significant variations. 

In this section, we focus on the overlay construction problem. We begin with defining the optimization goal, and 
present several scalable algorithms to construct an overlay. We then briefly discuss our approach to handling 
structural changes to the data graph.




\subsection{Preliminaries}
\label{problem}
As a first step, we convert the given data graph $\mathcal{G}(V, E)$ into an equivalent bipartite
graph $A_\calG(V', E')$, by identifying the query 
nodes, and the input nodes for each of the query nodes, given the user-provided query (as discussed in Section \ref{sec:model}).
We use the total number of edges in the overlay as our optimization metric, the intuition being that, each edge in the 
overlay corresponds to a distinct data movement and computation. We justify the use of this somewhat abstract metric by noting that
the runtime cost of an overlay is highly dependent on the distribution of the read/write frequencies; for the same query and
data graph, the optimal overlays could be wildly different for different distributions of read/write frequencies (which 
are not available at the overlay construction phase). We believe that the use of an abstract metric that rewards sharing is
likely to prove more robust in highly dynamic environments. In future work, we plan to further validate our choice by comparing
it against other optimization metrics.

More formally, we define 
the {\em sharing index} of an overlay to be:\\[-10pt]
$$ 1 - \frac{\text{\em \# of edges in the overlay}}{\text{\em \# of edges in }A_\calG}$$
\\[-8pt]
Figure \ref{fig:model-example} shows three overlays for our running example, and their sharing indexes. 
Figure~\ref{fig:model-example}(a) shows an overlay where
there are multiple paths between some reader-writer pairs. As we discussed earlier, such an 
overlay cannot be used for a duplicate-sensitive aggregate function (like \suma{}, \counta{}, etc.),
but for duplicate-insensitive aggregate functions like \maxa{}, it typically leads to better sharing
index as well as better overall throughput.
The second overlay uses  {\em negative edges} to bring down sharing index. 
This should only be done for aggregate functions where the subtraction operation is
incrementally computable (e.g., \suma{}, or \counta{}). 
Finally, third overlay is an example of a {\em
multi-level} overlay, and has the lowest sharing index for our running example (without use of 
negative edges or duplicate paths). In most cases, such multi-level overlays exhibit the best sharing index.
Note that multi-level overlays can also be duplicate insensitive or contain negative edges.

The problem of maximizing the sharing index is closely related to the {\em minimum order bi-clique
partition} problem~\cite{graphcompression}, where the goal is to cover all the edges in a bipartite graph
using fewest edge-disjoint bicliques. In essence, a biclique in the bipartite graph $A_{\calG}$
corresponds to a set of readers that all share a common set of writers. Such a biclique can thus 
be replaced by a partial aggregation node that aggregates the values from the common set of 
writers, and feeds them to the readers. In Figure \ref{fig:longexample}(d), node $PA_1$ corresponds
to such a biclique (between writers $a_w, b_w, c_w$ and readers $c_r, d_r, e_r, f_r, g_r$).
Finding bicliques is known to be NP-Hard. Sharing index (SI) is also closely related to the {\em
compression ratio} (CR) metric used in many of the works in {\em representational graph
compression}
~\cite{buehrer2008scalable}
; specifically, $CR = 1/(1 - SI)$. However, given 
the context of aggregation and the possibility of having {\em negative} 
and {\em duplicate-insensitive} edges in the overlay, 
we differentiate it from compression ratio. 
The problem of finding a good overlay is also closely related to the problem of {\em frequent
pattern mining}~\cite{graphmining3,fptree} as we discuss in the next section.

\subsection{Overlay Construction Algorithms}
\label{sec:ouralgo}

In this section, we present our algorithms for constructing different types of overlays as outlined
in the previous section. Given the NP-Hardness of the basic problem, and further 
the requirement to scale the algorithms to graphs containing tens of millions of nodes, we develop a
set of efficient heuristics to achieve our goal.
Our first set of proposed algorithms (called \vnma{}, \vnmn{}, and \vnmd{}) builds upon a prior algorithm (called \vnm{}) for bipartite graph compression 
by Buehrer et al.~\cite{buehrer2008scalable}, which itself is a adaptation of the well-known
FP-Tree algorithm for frequent pattern mining~\cite{graphmining3,fptree}. In our exploratory evaluation, we found
that algorithm to offer the best blend of scalability and adaptability for our purposes. 
Our second algorithm (called \iob{}) is an incremental algorithm that builds the overlay one reader at a time.

%

\subsubsection{Background: FP-Tree and VNM Algorithms}
\label{sec:bg}
\begin{figure}[t]
\centering
\hspace{-.2in}
 \includegraphics[width=88mm]{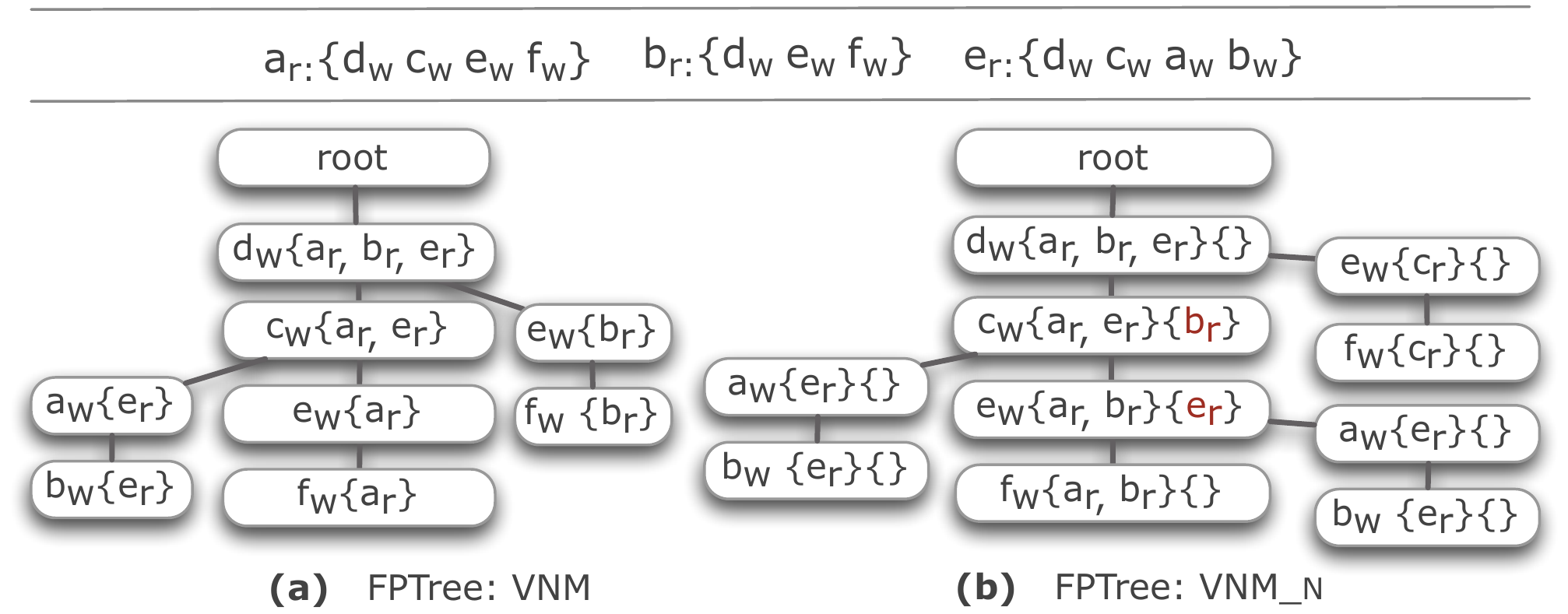}
 \caption{An example of FP-Tree construction for VNM and VNM$_N$: (a) Basic version, (b)
FP-Tree with negative edges.}
\label{fig:FP-Tree}
\end{figure}

We begin with a brief recap of the {\em FP-Tree} algorithm for frequent pattern mining,
   considered to be one of the most efficient and scalable algorithms for finding frequent patterns. 
%
We briefly outline the algorithm using the terminology of readers and writers rather than
transactions and items. 
First, the writers are sorted in the increasing order by their overall frequency of occurrence in
the reader input sets, i.e., their out-degree in $A_\calG$. In our running example, the sort order
(breaking ties arbitrarily) would be $\{d_w, c_w, e_w, f_w, a_w, b_w\}$. Then all the reader input
lists are rewritten according to that sort order; e.g., we would write the input list of $a_r$
as $\{d_w, c_w, e_w, f_w\}$. Next, the FP-Tree is built incrementally by adding one reader at a time, starting with 
an empty tree. For the reader under consideration, the goal is to find its longest prefix that
matches with a path from the root in the FP-Tree constructed so far. As an example,
        Figure~\ref{fig:FP-Tree} shows the FP-Tree
built after addition of readers $a_r$, $b_r$, and $e_r$. A node in the FP-Tree is represented by:
$x_w \{S(x_w)\}$ where $x_w$ is a writer and $S(x_w)$ is a list of readers that contain $x_w$ in their
input lists (called {\em support set}). Now, for reader $c_r$, the longest prefix of it that matches
a path from root is $d_w, c_w, e_w, f_w$. That reader would then be added to the tree nodes in that path 
(i.e., to the support sets along that path). 
If the reader input list contains any additional writers, then a new branch is created in the tree (for $e_r$ a new
        branch will be created with nodes $a_w\{e_r\}$ and $b_w\{e_r\})$.

Once the tree is built, in the {\em mining phase}, the tree is searched to find
bicliques. A path $P$ in the tree from the root to the node $x_w\{S(x_w)\}$ corresponds 
to a biclique between the writers corresponding to the nodes in $P$ and the readers in $S(x_w)$.
Since our goal is to maximize the number of edges removed from the overlay graph, we search
for the biclique that maximizes: 
\\[2pt] {\hspace*{10mm}$benefit(P) = L(P)*|S(P)| - L(P) - |S(P)|, $}\\[2pt]
where $L(P)$ denotes the length
of the path $P$ and $S(P)$ denotes the support for the last node in the path. Such a biclique can
be found in time linear to the size of the FP-Tree.
%
After finding each such biclique, ideally we should remove the corresponding edges (called the {\em mined edges}) and reconstruct 
the FP-Tree to find the next biclique with best benefit. Mining the same FP-Tree would still find bicliques but  with lower
benefit (since the next biclique we find cannot use any of the edges in the previously-output biclique). 

We now briefly describe the \vnm{} algorithm~\cite{buehrer2008scalable}, which is a highly scalable
adaptation of the basic FP-Tree mining approach described above; \vnm{} was developed for
compressing very large (web-scale) graphs, and in essence, replaces each biclique with a virtual
node to reduce the total number of edges. 
The main optimization of \vnm{} relies on limiting the search space by creating small groups of readers, and looking for 
bicliques that only involve the readers in one of the groups. This approach is much more scalable than building an
FP-Tree on the entire data graph. \vnm{} uses a heuristic based on {\em shingles}~\cite{shinglesravi,shinglesmotwani} 
 to group the readers. 
Shingle of a reader is effectively a signature of its input {\em writers}. 
If two {\em readers} have very similar {\em adjacency lists}, then with high probability, their
shingle values will also be the same. In a sense, grouping readers by shingles increases the chance of finding big bicliques (with higher benefit) within the groups. 
The algorithm starts by computing multiple shingles for each reader, and then doing a lexicographical sort of the
readers based on the shingles. The sorted list is then chunked into equal-sized groups of readers, each of which is passed
to the FP-Tree algorithm separately. 
Mining all the reader groups once completes one iteration of the algorithm. The process is then repeated 
with the modified bipartite graph (where each biclique is replaced with a virtual node) to further compress
the graph. Since the virtual nodes are treated as normal nodes in such subsequent iterations, a biclique 
containing virtual nodes may be replaced with another virtual node, resulting in connections between virtual nodes; in our context, this gives rise to 
{\em multi-level overlays} where partial aggregation nodes feed into other partial aggregators.

\subsubsection{$VNM_A$: VNM Adaptive}
Our first adaptation of the basic \vnm{} algorithm is aimed at addressing a major deficiency of that algorithm, namely lack of a 
systematic way to choose the chunk size. 
Our initial experiments with \vnm{} suggested that the effect of the chunk size on the final
compression achieved is highly non-uniform across various graphs like web graphs and social graphs.
We noticed that a bigger chunk size typically finds bigger bicliques, but it can't find all big
bicliques, especially when there is big overlap in the reader sets of two potential bicliques. This
is because the reader sets of two subsequent mining phases in \vnm{} are mutually exclusive. 
Second, a bigger chunk size makes it harder to find small bicliques, which is especially a problem
with later iterations; since many of the original edges have been deleted in the first few
iterations, only small bicliques remain in the graph. On the other hand, using a small chunk size
from the beginning ignores large bicliques that can deliver huge savings. 

To address this problem, we develop an adaptive variation of \vnm{} that uses different chunk sizes
for different iterations. 
For the first iteration, we use a large chunk size (100 in our experiments) and dynamically reduce it
for future iterations. For the $i^{th}$ iteration, let $c_i$ denote the chunk size, and let $B_i^s$
denote the sum total of the {\em benefits} (defined in Section~\ref{sec:bg}) for all the bicliques found in that iteration with reader
set size $= s$ (note that, $s \le c_i$). We choose $c_{i+1} \le c_{i}$ to be the smallest $c$ such that:
$\sum_{s \le c} B_i^s > 0.9 \sum_{s \le c_{i}} B_i^s$. 
Although our algorithm also requires setting two parameter values, our extensive experimental evaluation on 
many real-world graphs showed that the algorithm is not sensitive to the initial chunk size to within
an order of magnitude, and to the second parameter between 0.8 and 1.

\subsubsection{$VNM_N$: VNM with Negative Edges}
Next, we present our adaptation to \vnm{} that considers adding negative edges to reduce the overlay size.
In essence, we look for {\em quasi-bicliques} that may be a few edges short of being complete bicliques
(this problem is also known to be NP-Hard~\cite{liu2008quasi}).
For scalability, our algorithm employs the same basic structure as the \vnma{} algorithm discussed above (with grouping of 
readers using shingles); however, we modify 
the FP-Tree construction and mining algorithms to support negative edges.

Recall that a node in an FP-Tree is represented by $x_w\{S(x_w)\}$
where $x_w$ is a writer and $S(x_w)$ contains the readers that contain $x_w$ in their
input lists. To accommodate negative edges, we now represent a node by $x_w\{S(x_w)\}\{S'(x_w)\}$,
where $S'(x_w)$ contains readers that do not contain $x_w$ in their input list, but
may contain the writers corresponding to the nodes below this node in the FP-Tree. Benefit of a 
path $P$ in the FP-Tree is now given by: \\[2pt]{\hspace*{2.5mm} \small $benefit(P) = L(P)*|S(P)| - L(P) -  |S(P)| - \sum_P
|S'(x_w)|$},  \\[2pt]
where the last term captures the number of negative edges along $P$. 

In our proposed algorithm, when an FP-Tree is augmented to include a new reader $r$, we add $r$
along up to $k_1$ paths in the FP-Tree that maximize the benefit given the FP-Tree constructed so far.
More specifically, we exhaustively explore the FP-Tree in a breadth-first manner, and for each node
visited, we compute the benefit of adding $r$ along the path. We then choose
up to $k_1$ paths with the highest benefit and add the reader along those paths. As with
the original FP-Tree algorithm, additional branches may have to be created for the remaining writer
nodes in $r$. 

Figure~\ref{fig:FP-Tree}(b) shows an example where upto two paths can be added for a reader (i.e., $k_1 = 2$). 
Both readers $b_r$ and $e_r$ create two paths in the overlay, one of which uses a negative edge. 
Note that $e_r$ creates a new branch in the tree
(apart from the one similar to in the basic version); after the introduction of negative edge for $e_r$ at $a_w$, there are
still nodes remaining in $e_r$'s input list.
During the mining phase the new FP-Tree finds a biclique of size {\em 3x3}. On
the other hand, the basic version can only find a biclique of size {\em 2x2}.


Although our algorithm finds the best paths to add the reader along, it runs in time linear to the
size of the FP-Tree constructed so far. However, since the FP-Tree, in essence, now encodes information about
$k_1$ times as many readers as it did before, the size of the FP-Tree itself is expected to be 
larger by about the same factor. To improve efficiency, we stop the breadth-first exploration down a
path if more than $k_2$ negative edges are needed to add $r$ along that path (we set $k_2 = 5$ in our
experiments). This optimization has little impact on performance since it is unlikely
that quasi-bicliques requiring a large number of negative edges will be beneficial.

\subsubsection{$VNM_D$: Duplicate-insensitive VNM}
Next, we discuss our proposed algorithm for finding overlays that exploit the duplicate-insensitive
nature of some aggregates and allow for multiple paths between a writer and a reader. 
There are two natural ways to extend the \vnm{} algorithm for reusing edges in this fashion. First, we can keep the basic structure of the 
\vnm{} algorithm and modify the FP-Tree algorithm itself to find multiple bicliques in each mining
phase, while ignoring the overlap between bicliques. However, by construction, the bicliques mined
from a single FP-Tree tend to have very high overlap, and the benefits for additional bicliques
found can be very low. It is also not clear how many aggregate nodes to add in a single mining
phase; adding all bicliques for which the benefit is non-zero is likely to lead to many partial
aggregate nodes, each providing low benefit.

Instead, in our proposed algorithm \vnmd{}, we modify the reader grouping phase itself. In \vnm{}, in each
iteration, the readers are grouped into disjoint groups before passing to the FP-Tree construction
and mining phase. Instead, we allow the groups of readers to overlap. Specifically, given an overlap percentage $p$ 
(an algorithm parameter), we allow two consecutive groups of readers to have $p\%$
readers in common. The FP-Tree construction and
mining phases themselves are unchanged with the following exceptions. First, instead of representing an
FP-Tree node as $x_w\{S(x_w)\}$, we represent it as $ x_w\{S^{{not mined}}(x_w)\}\{S^{{mined}}(x_w)\}$, where
$S^{{mined}}$ $(x_w)$ contains the readers $r$ such that the edge from
$x_w$ to $r$ was present in a previously used biclique. 
Second, we modify the formula for computing the benefit of a path as
follows: \\[2pt]
{\small $benefit(P) = L(P)*|S(P)| - L(P) - |S(P)| - \sum_P |S^{mined}(x_w)|$}; \\[2pt]
the last term captures the number of {\em reused} edges in the biclique. %

\subsubsection{IOB: Incremental Overlay Building}
The overlay constructions algorithms that we have developed so far are all based on identifying
sharing opportunities by looking for bicliques in $A_\calG$. However, to make those algorithms scalable,
two heuristics have to be used: one to partition the readers into small groups, and one to mine the
bicliques themselves. In essence, both of these focus the search for sharing opportunities to
small groups of readers and writers, and never consider the entire $A_\calG$ at once. Next, we present an
incremental algorithm for building the overlay that starts with an empty overlay, and adds one reader 
at a time to the overlay. For each reader, we examine the entire overlay constructed till that point which,
as our experimental evaluation demonstrates, leads to more compact overlays.

%
%


We begin with ordering the readers using the shingle order as before, and add the readers one at a
time in that order. In the beginning, the overlay graph simply contains the (singleton) writer nodes.
Let $\langle r, \calN(r)\rangle$ denote the next reader to be added. 
Let $\langle ovl_n, \calI(ovl_n) \rangle$ denote a node in the overlay constructed so far, where
$\calI(ovl_n)$ is the set of writers whose partial aggregate $ovl_n$ is computing. For reader $r$,
our goal is to reuse as much of the partial aggregation as possible in the overlay constructed so
far. 
In other words, we would like to find the smallest set of overlay nodes whose aggregates can be used
to compute the aggregate for $r$.
This problem is essentially the {\em minimum exact set cover problem}, which is known to be
NP-Complete.

We use a standard greedy heuristic commonly used for solving the set cover problem.  We start by finding the overlay node that
has maximum overlap with $\calN(r)$, and restructure the overlay to make use of that overlap. We
keep on repeating the same process until all nodes in $\calN(r)$ are covered (since the singleton
writer nodes are also considered part of the overlay, we can always cover all the nodes in
$\calN(r)$). 
Let $\langle v_1, B\rangle$ denote
the overlay node that was found to have the highest overlap with the uncovered part, denoted $A$, of
$\calN(r)$. If $B \subseteq A$, then we add an edge from $v_1$ to $r$, and repeat the process with
$A - B$. Otherwise, we restructure the overlay to add a new node $\langle v_1', A \cap B
\rangle$, reroute the appropriate incoming edges (i.e., the incoming edges corresponding to the
writers in $A \cap B$) from $v_1$ to $v_1'$, and add a directed edge from $v_1'$ to
$v_1$. We then also add an edge from $v_1'$ to $r$. If $A - A \cap B$ is non-empty,  then we repeat
the process to cover the remaining inputs to $r$.

As with the \vnm-based algorithms, we use multiple iterations to improve the overlay. In each iteration (except the 1st iteration), we
revisit the decisions made for each of the partial aggregator nodes, and do local restructuring of the overlay if better 
decisions are found for any of the partial aggregator nodes (using the same set cover-based algorithm as above).

\begin{figure*}[t]
\centering
 \includegraphics[width=125mm]{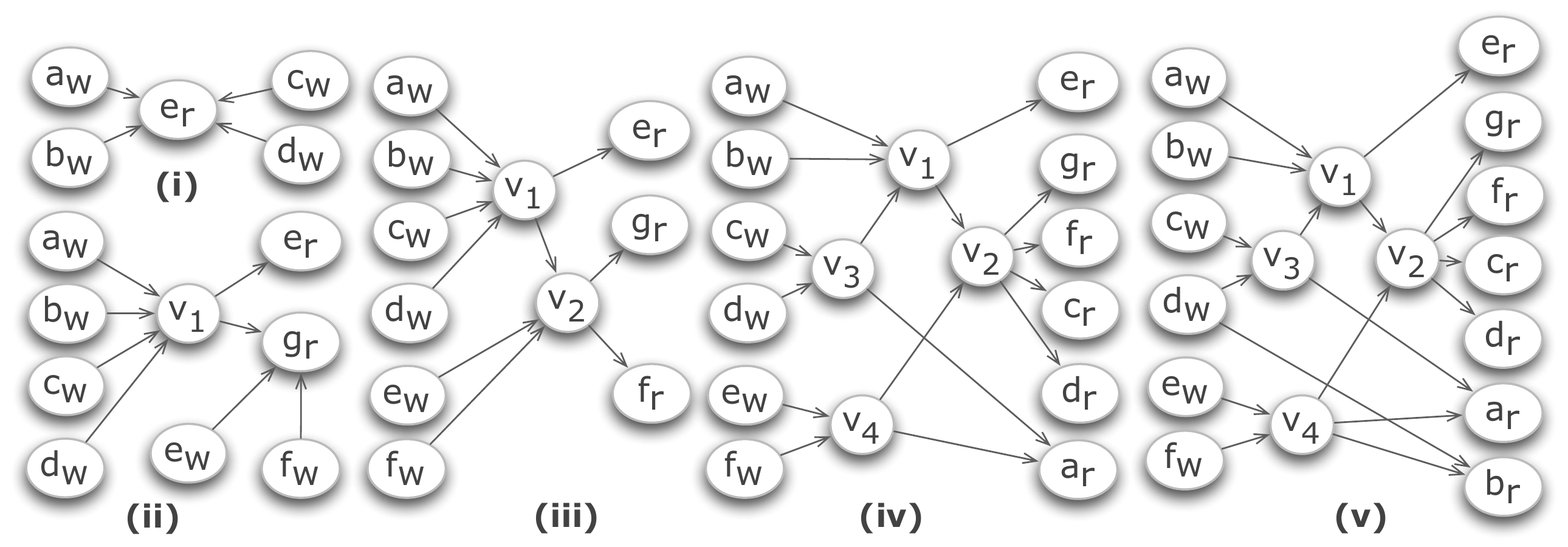}
 \caption{An example execution of \iob{} algorithm for the data graph in Figure~\ref{fig:longexample}(a). }
\label{fig:IOB_example}
\end{figure*}

\topicu{Example}
Figure~\ref{fig:IOB_example} shows an overlay built by \iob{} for our running example, with the readers added in order:
$\{e_r,g_r,f_r,c_r,d_r,a_r,b_r\}$. Figure~\ref{fig:IOB_example}(i) shows the overlay after adding the
$\langle e_r, \{a_w, b_w, c_w, d_w\}\rangle$. While adding the second reader, i.e., $\langle g_r, \{a_w, b_w, c_w,
d_w, e_w, f_w\}\rangle$, $\calI(e_r)$ is found to have maximum overlap with 
$\calN{(g_r)}$. Hence, a new aggregator node $v_1$ (aggregating $a_w,
b_w, c_w, d_w$) is added and shared by $e_r$ and $g_r$ (Figure~\ref{fig:IOB_example}(ii))\footnote{\scriptsize We do not
allow a reader node to directly form an input to an aggregator node.}. Also direct edges are
added from $e_w$ and $f_w$ to $g_r$ (for the inputs that were not covered). 
For $f_r$: $\calI(g_r)$ is found to have the maximum overlap with $\calN(f_r)$, leading to the construction of a new
aggregator node $v_2$, and so on. We omit a detailed discussion of the rest of the construction due to space constraints.

\vspace{8pt}
For efficient execution of the algorithm we maintain both a {\em reverse index} and a {\em forward index}. For a writer node
$w$, the reverse index tells us which are the overlay nodes that are aggregating $w$. For example, $a_w$'s  reverse index
entry
will have both $v_1$ and $v_2$. Note that  even though there is no direct edge from $a_w$ to $v_2$,  $a_w$'s reverse
    index entry has $v_2$ because  $v_2$ is effectively aggregating $a_w$. This index helps us to find the overlay node that
    provides maximum cover to a set of input nodes using one single scan of the input list. On the other hand, for any node $n$ in
    the overlay, the forward index tells us the input list of $n$. For example, $v_2$'s forward index entry will have $v_1$
    and $v_3$ in it. 

Although the above algorithm could be extended to allow for negative edges and/or duplicate paths,
we do not discuss those extensions here. This is because, although \iob{} finds significantly smaller
overlays, the overlays tend to be deep (with many levels) and in our experimental evaluation, the
end-to-end throughput for the overlays was lower than for the overlays found with the \vnma{}
algorithm. Thus, although the \iob{} algorithm is a better algorithm for finding compact overlays
and for compressing bipartite graphs, \vnm-based algorithms are better suited for our purpose.

\subsection{Handling Dynamic Changes}
\label{sec:hdc}
\eat{
We adapt the basic ideas underlying the IOB algorithm to incrementally update the overlay in response to 
structure changes to the underlying graph. Our system can handle both addition and deletion of edges, 
as well as addition and deletion of nodes. Due to space constraints, we omit that discussion here and
refer the reader to the extended version of the paper.
}

We briefly sketch our techniques to incrementally update the overlay in response to structural
changes to the underlying data graph.


\topicu{Addition of Edges} When a new edge is added to the data graph, we explore the neighborhoods
of both the endpoints to construct a list: $\{ \langle r_1, \Delta(\calI(r_1)) \rangle, \langle r_2, \Delta(\calI(r_2))
    \rangle, ... \}$, where $r_i$ denotes a reader whose input list has changed, and 
$\Delta(\calI(r_i))$ denotes the new writer nodes that are added to $\calI(r_i)$ (for queries
defined over 2-hop neighborhoods or larger, the changes to the input lists can be substantial). 
We process the change for each of the readers separately. If $|\Delta(\calI(r_i))|$ is larger than a
prespecified {\em threshold}, we
use the \iob{} algorithm to add a new aggregate node that computes a partial aggregate over the writers
in $\Delta(\calI(r_i))$ (in the best
case scenario, an existing overlay node may already compute exactly this aggregate); we then add an edge
from that node to $r_i$.
Otherwise, we add direct edges from
the writer nodes in $\Delta(\calI(r_i))$ to $r_i$. At the same time, for each reader in the overlay
$r$, we also keep a count of direct edges from writers to that reader; this count is updated in the
latter case, and if the count is larger than the {\em threshold}, we use the \iob{} algorithm to
restructure the overlay as above.



\topicu{Deletion of Edges} Edge deletions are trickier to handle because significant restructuring of
the overlay may be needed,
especially for multi-level overlays and complex ego-centric
queries. As above, we explore the neighborhoods of the endpoints of the deleted edge to construct a list: $\{ \langle
r_1, \Delta(\calI(r_1)) \rangle, \langle r_2, \Delta(\calI(r_2)) \rangle, ... \}$,
and we process each reader independently. For each of the readers, we make a pre-processing pass
over the overlay to identify how many of the overlay nodes would need to be modified to accommodate
the change. In other words, for reader $r_i$, we count the number of overlay nodes $ovl$ that are
upstream of $r_i$ and $\calI(ovl) \cap \Delta(\calI(r_i)) \ne \phi$. If this number is small 
($\le 5$), we modify the overlay by splitting the overlay nodes and removing edges appropriately (we
omit the details here for brevity). Otherwise, we simply remove all incoming edges to $r_i$ (and any 
partial aggregate nodes that only send values to $r_i$), and use the \iob{} algorithm to add $r_i$ back
in with the modified input list.


%
\topicu{Addition or Deletion of Nodes} Addition of a new node, $u$, to the data graph is easy to handle. 
First, we add a new writer node $u_w$ to the overlay, and then add direct edges from $u_w$ to
some of the existing reader nodes as dictated by the new edges added between $u$ and the existing
nodes (in most cases, a new node is added with one edge to an existing node). Second, we construct
an input list for $u$, and we use the \iob{} algorithm to add a new reader node $u_r$ with that input
list. 
On the other hand, when a node $u$ is deleted from the data graph, we simply remove $u_w$ and $u_r$
and all their incident edges from the overlay graph. We also remove $u$ from the data structures
used for incremental maintenance (i.e., the forward and the reverse indexes).

\begin{figure*}[th]
\centering
 \includegraphics[width=165mm]{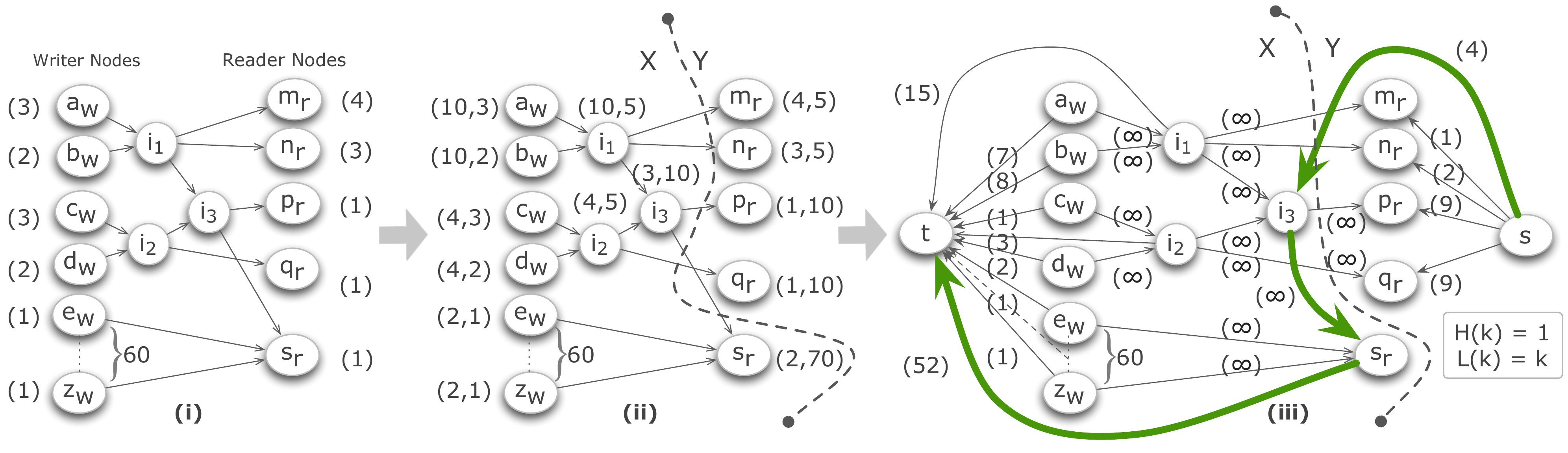}
 \caption{ (i) An example overlay annotated with read/write frequencies; (ii) Computing ({\em pull}, {\em push}) frequencies; (iii) Construction of the {\em s-t} augmented graph (with the annotations
         denoting the edge weights).}
\label{fig:examplepushpull}
\end{figure*}


\section{Making Dataflow Decisions}
\label{sec:pushpull}
Next, we discuss how to make the dataflow (i.e., precomputation) decisions to maximize the total throughput given an overlay network,
and the expected read/write frequencies for the nodes. Surprisingly, the problem can be solved
optimally in polynomial time. 
We begin with the preliminaries related to the cost of a dataflow decisions and then provide the formal problem definition and present the analysis along with the algorithms that we propose. 

\label{sec:pushpullproblem}
\subsection{Preliminaries} For each node $v \in V$ in the data graph, let $r(v)$ denote 
its read frequency (i.e., the number of times a query is issued at node $v$), let $w(v)$ denote
its write frequency (i.e., the number of times $v$ is updated)\footnote{\scriptsize See Table~\ref{table1} for a summary of notation.}. Given these, with each node $u \in V''$ in the overlay
graph $O_\calG(V'',E'')$, we associate two numbers, $f_l(u)$ and $f_h(u)$, called {\em pull frequency} and {\em
push frequency}, respectively. 
$f_h(u)$ captures the number of times
data values would be {\em pushed} to $u$ if all nodes in the overlay are assigned {\em push}
decisions. Similarly, $f_l(u)$ indicates the number of times data values would be {\em pulled} from
$u$ if all nodes in the overlay are assigned {\em pull} decisions. 

The push and pull frequencies are computed as follows. For computing push frequencies, we start by
assigning $f_h(a_w) = w(a_w)$ for all writer nodes $a_w$, and then propagate the push frequencies 
from left to right (downstream). For an aggregation node or a reader node
$u$, $f_h(u)$ is computed by summing up the push frequencies for all nodes that are immediately 
upstream of $u$. Similarly, the pull frequencies are computed by starting with the reader nodes, then recursively computing the pull frequencies for the rest of the nodes.
Figure~\ref{fig:examplepushpull}(i)-(iii) illustrates this with an example that we also use to show how our algorithm
makes the dataflow decisions.


\subsection{Push and Pull Costs} As discussed before, a {\em push} decision on a node implies that the aggregate corresponding to
that node will be (incrementally) precomputed and will be available for immediate consumption. On the other hand, a
{\em pull} decision on a node implies that the aggregate will be computed on demand when the node
is read. In order to reason about the tradeoff between push and pull, we need to be able to compute
the cost of a push or a pull. This cost typically depends on the nature of the aggregate, and the type
and the size of the sliding window~\cite{arasu2004resource}.
We capture these costs as two functions: $H(k)$ denotes the average cost of one push
for an aggregation node with $k$ inputs, and $L(k)$ denotes the average cost of one 
pull for that node.  For example, for a \suma{} aggregate node, we expect $H(k) \propto 1$ and $L(k) \propto k$, 
whereas for a \maxa{} aggregate node, if we use a priority queue for handling incremental updates, then 
$H(k) \propto \log_2(k)$ and $L(k) \propto k$. To handle sliding windows, we implicitly assign $w$ inputs to each writer 
where $w$ is the average number of values in the sliding window at a writer -- thus if the sliding window is of size 10, then $PUSH$ and 
$PULL$ costs of the writer node will be $H(10)$ and $L(10)$ respectively. 
We assume $H()$ and $L()$ are either provided, or are computed through a calibration
process where we invoke the aggregation function for a range of different inputs and 
learn the $H()$ and $L()$ functions. 

\subsection{Problem Definition} The dataflow decisions made by a solution induce a {\em node partition}, denoted $(X, Y), X \cap Y = \phi$, on the overlay graph, 
where $X$ contains nodes that are designated {\em push}, $Y$ contains nodes designated {\em pull} (Figure~\ref{fig:examplepushpull}(ii)). 
Since all nodes upstream of a {\em push} node must also be designated {\em push} (and similarly all nodes downstream
of a {\em pull} node must also be {\em pull}), the partition induced by any consistent 
set of dataflow decisions must satisfy the constraint that there is no edge from a node in $Y$ to a node in $X$. 

For an overlay node $v$, let {$PUSH(v)$} $= f_h(v)*H(deg(v))$ denote the cost incurred if it is designated a
{\em push} node, let {$PULL(v)$} $= f_l(v)*L(deg(v))$ denote the cost if it is a {\em pull} node.
Although the push/pull decisions cannot be made for the nodes independently (because of the aforementioned constraint), 
$PUSH()$ and $PULL()$ costs can be computed independently; this is because the computations that happen at a node when 
it is invoked, do not depend on the dataflow decisions at its input or output nodes. Thus, 
to minimize the total computational cost, our goal reduces to finding an ($X$, $Y$) partition of the overlay (with no edges going from 
$Y$ to $X$) that minimizes:
{$\sum_{v \in X} PUSH(v) + \sum_{v \in Y} PULL(v)$}.

\topicu{Query Latencies} Another consideration in making dataflow decision is the impact on query {\em latencies}. Throughput maximization 
may lead to higher use of {\em pull} decisions, especially if reads are less frequent than writes, that may result 
in high query latencies. As we show in Section \ref{sec:evaluations}, because our system is entirely in-memory and does not need to do distributed network
traversals, the query latencies are quite low even in the worst-case. In future work, we plan to investigate latency-constrained
optimization as well as understand the interplay between throughput and latency better.


\label{label:maxflow}

\subsection{Algorithm}
We design an algorithm for a slightly more general problem that we call {\em difference-maximizing partition} problem,
that we describe first. 
    
\topicu{Difference-Maximizing Partition (DMP) Problem}
We are given a directed acyclic graph $H(H_V, H_E)$, where each vertex $v \in H_V$ is associated with a weight $w(v)$;
$w(v)$ may be negative. For ease of exposition, we assume that $\forall v, w(v) \ne 0$. We are asked to find a graph partition $(X, Y)$, such that
there are no edges from $Y$ to $X$, that maximizes: $\sum_{v \in X} w(v) - \sum_{v \in Y} w(v)$. Note that, the solution
is trivially $(X = H_V, Y = \phi)$ if all node weights are positive. We also note that, the metric has the maximum
possible value if all nodes with $w(v) < 0$ are assigned to $Y$, all nodes with $w(v) > 0$ to $X$. However, that particular
assignment may not guarantee that there are no edges from a vertex in $Y$ to a vertex in $X$.


\topicu{Reducing Dataflow Decisions Problem to DMP}
To reduce our problem to this problem, we set: \\ 

\vspace{-6pt}
{\small
$w(v) = f_l(v) L(deg(v)) - f_h(v) H(deg(v))$} = {\scriptsize $PULL(v) - PUSH(v)$} \\ 

\vspace{-6pt}
\noindent 
That is,
the weight of node $v$ is the ``{\em benefit}'' of assigning it a {\em push} decision (which is negative if $PULL(v) < PUSH(v)$).
Then:

\vspace{-5pt}
{
\scriptsize
\begin{align*}
&\sum_{v \in X} w(v) - \sum_{v \in Y} w(v)  \\
&=  \sum_{v \in X} (PULL(V)- PUSH(v)) - \sum_{v \in Y} (PULL(V) - PUSH(v)) \\
&= \underline{\sum_{v \in H_V} (PULL(v) + PUSH(v))} - 2 (\sum_{v \in X} PUSH(v)  + \sum_{v \in Y} PULL(v))
\end{align*}
}
\noindent
Since the underlined term is a constant, maximizing this is equivalent to minimizing {\small $\sum_{v \in X} PUSH(v)  + \sum_{v \in Y} PULL(v)$}.

\eat{
$$\sigma_{v \in S} w(v) - \sigma_{v \in T} w(v) = $$
$$\sigma_{v \in S} (f_l(v) L(deg(v)) - f_h(v) H(deg(v)) ) - \sigma_{v \in T} (f_l(v) L(deg(v)) - f_h(v) H(deg(v))) = $$
$$\sigma_{v \in H_V}   (f_l(v) L(deg(v)) + f_h(v) H(deg(v))) - 2 \times (\sigma_{v \in S} f_h(v) H(deg(v)) + \sigma_{v \in T} (f_l(v) L(deg(v)))) $$
}

\topicu{Algorithm for Solving DMP}
To solve this more general problem, we construct an edge-weighted graph $H'(H'_V, H'_E)$ from $H(H_V, H_E)$ (in practice, we do not make a copy
but rather augment $H$ in place). $H'_V$ contains all 
the vertices in $H_V$ and in addition, it contains a {\em source} node $s$ and a {\em sink} node $t$ (nodes in $H'$ are unweighted). 
Similarly, $H'_E$ contains all the edges in $H_E$, with edge weights set to $\infty$.
Further, for each $v \in H_V$ such that $w(v) < 0$, we add a directed edge in $H'$ from $s$ to $v$ with weight $w'(s,v) = -w(v)$. Similarly for $v \in H_V, s.t.\ w(v) > 0$,
we add a directed edge in $H'$ from $v$ to $t$ with weight $w'(v, t) = w(v)$ 
(see Figure~\ref{fig:examplepushpull}(iii) for an example). 

We note that, this construction may seem highly counter-intuitive, since a lot of nodes in $H'$ have either
no outgoing or no incoming edges and there are few, if any, directed paths from $s$ to $t$. In fact,
the best case scenario for the algorithm is that: {\em there is no directed path from $s$ to $t$}.
This is because, a path from $s$ to $t$ indicates a {\em conflict} between two or more nodes.
The highlighted 
path form $s$ to $t$ in Figure~\ref{fig:examplepushpull}(iii) provides an example. 
The best decision for node $i_3$ in isolation would be {\em pull} ($PULL(i_3) = 6, PUSH(i_3) = 10$), but
that for $s_r$ is {\em push} because of its high in-degree and because $L(k) = k$ ($PULL(s_r) = 2*60 = 120, PUSH(s_r) = 70$).
However, a {\em pull} on $i_3$ would force a {\em pull} on $s_r$, hence both of them cannot be assigned the optimal decision in isolation.

After constructing $H'$, we find an $s$-$t$ directed min-cut in this directed graph, i.e., a set of edges $C \in H'_E$ with minimum total edge-weight, such that removing those edges leaves no 
directed path from $s$ to $t$. 
Let $Y$ denote the set of nodes in $H'$ reachable from $s$ after removing the edges in $C$ (excluding $s$), let $X$ denote the set of remaining nodes 
in $H'$ (excluding $t$).

\vskip 6pt
\begin{theorem}
$(X, Y)$ 
is a node partition of $H$ such that there are no edges from $Y$ to $X$, and $\sum_{v \in X} w(v) - \sum_{v \in Y} w(v)$ is maximized. 
\end{theorem}
\vskip 6pt

\noindent{\bf Proof:}
We prove that the original problem of finding an $(X, Y)$ partition of $H$ with the desired properties is equivalent to finding a $s$-$t$ min-cut 
in $H'$. 

\begin{figure}[th]
\centering
 \includegraphics[width=80mm]{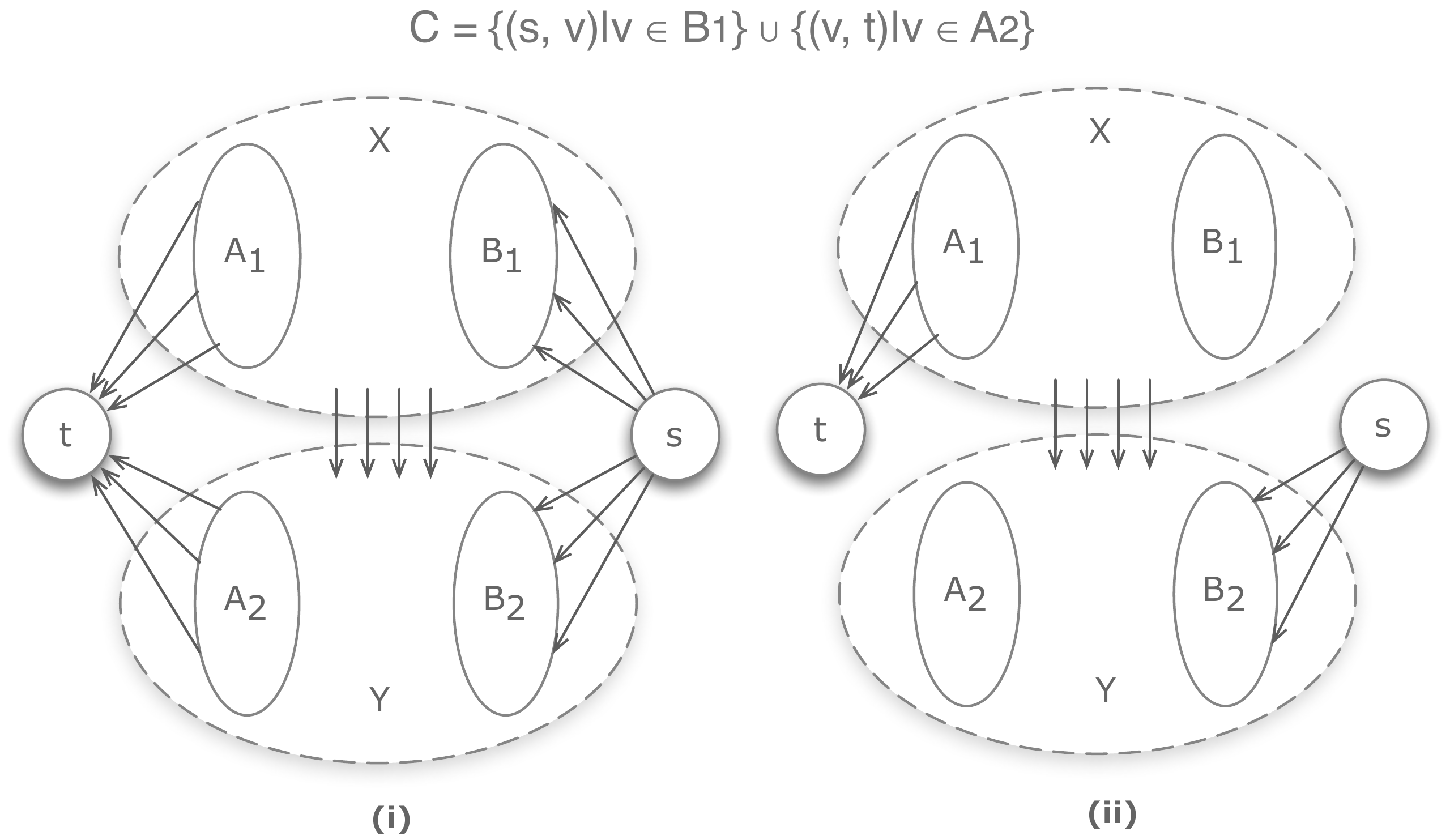}
 \caption{(i) Structure of an optimal solution, (ii) Removing C leaves no path between {\em s} and {\em t}. }
\label{fig:proof}

\end{figure}

Let $(X, Y)$ denote an optimal solution to our original problem, i.e., a partition of $H(H_V, H_E)$  such that the edges are directed from $X$ to $Y$ and 
$\sum_{v \in X} w(v) - \sum_{v \in Y} w(v)$ is maximized. 
Let $B = \{v \in H_V | w(v) < 0\}$ denote the vertices that have an edge from
$s$ in $H'$, and let $A = \{v \in H_V | w(v) > 0\}$ denote the vertices that have an edge to $t$ in $H'$. Figure \ref{fig:proof} shows the 
structure of the optimal solution, where we define $A_1 = A \cap X, B_1 = B \cap X, A_2 = A \cap Y,$ and $B_2 = B \cap Y$.

Let $C = \{ (s, v) | v \in B_1 \} \cup \{ (v, t) | v \in A_2 \}$. Since $(X, Y)$ is a partition of $H$ such that there are no (directed) edges from $Y$ to $X$,
removing $C$ from $H'$ leaves no path from $s$ to $t$ (although there are edges from $s$ to nodes in $B_2$ and from nodes in $A_1$ to $t$, there can be no path from a node in $B_2$ to a node in $A_1$ since
$B_2 \subseteq Y$ and $A_1 \subseteq X$). In other words, $C$ is an $s$-$t$ directed cut. Now: 

{
\begin{align*}
 \sum_{v \in X} & w(v) -  \sum_{v \in Y} w(v) \\
&=  \sum_{v \in A_1} w(v) + \sum_{v \in B_1} w(v) - \sum_{v \in A_2} w(v) - \sum_{v \in B_2} w(v) \\
&= \underline{\sum_{v \in A} w(v) - \sum_{v \in B} w(v)} - 2 (\sum_{v \in B_1} (-w(v)) + \sum_{v \in A_2} w(v) ) \\
&= \underline{\sum_{v \in A} w(v) - \sum_{v \in B} w(v)} - 2 (\sum_{e \in C} w'(e))
\end{align*}
}

\noindent
Since the underlined term is a constant, the optimal solution is such that $\sum_{e \in C} w'(e)$
is minimized, i.e., $C$ is an $s$-$t$ directed min-cut with the constraint that there is no edge from $Y$ to $X$.

Thus the last thing we need to prove is that: if $C$ is an $s$-$t$ directed min-cut of $H'$, then the corresponding partition $(X, Y)$
of $H_V$ satisfies this constraint. 

First we note two things. First, $C$ cannot contain any of the edges from the original edges $H_E$ (that all have weight $\infty$).
This is because, the set of all outgoing edges from $s$ is a valid $s$-$t$ cut that has a finite value, so a min-cut must have
finite value. Second, given any $s$-$t$ cut $C$ that does not include any of the original edges in $H_E$, we can 
define a corresponding partition $(X, Y)$ uniquely, where $Y$ contains nodes that are reachable from 
$s$ (excluding $s$), and $X$ contains the remaining nodes in $H'_V$ (excluding $t$).

Assume to the contrary that, in the $(X, Y)$ partition defined by the min-cut $C$, there is a directed edge $(u, v)$ such that $u \in Y$ and $v \in X$. Since 
the weight of this edge is $\infty$, it will not be part of the cut. There are four cases:
\begin{itemize}
\item $u \in B_2, v \in A_1$: Since $(s, u)$ and $(v, t)$ are $\notin C$, we get a path from $s$ to $t$, contradicting the assumption
that $C$ is a cut.
\item $u \in B_2, v \in B_1$: This means there is a path from $s$ to $v$ (through $u$) even after deleting the edges in $C$, and thus $(s, v)$ 
can be removed from $C$ without compromising the property that $C$ is an $s$-$t$ cut. In other words, $C$ is not a min-cut.
\item $u \in A_2, v \in B_1$: Note that there must be a path from $s$ to $u$ after removing $C$, otherwise we wouldn't need to include
$(u, t)$ in the cut. Then, as above, we have a path from $s$ to $v$, and $C$ is not a min-cut.
\item $u \in A_2, v \in A_1$: As above, there is a path from $s$ to $u$ after removing $C$, and combined with $(v, t)$, we get a path from
$s$ to $t$ contradicting the assumption that $C$ is a cut.
\end{itemize}

We thus conclude that, finding an optimal $(X, Y)$ partition of $H$ that minimizes the 
objective function is equivalent to finding a $s$-$t$ directed min-cut in $H'$.

\noindent
$\square$


We use the Ford-Fulkerson algorithm to construct an $s$-$t$ max-flow in $H'$, and use it to find the optimal $(X, Y)$ partition of $H$.
Specifically, after the max-flow solution has been obtained, we perform a breadth-first search from $s$ to find all nodes reachable from it in the
residual graph of the max-flow, assign those nodes to $Y$, and assign the rest of the nodes to $X$.

%
%

\vspace{10pt}
\subsection{Pre-processing} Although the above algorithm runs in polynomial time, it is not feasible to run max-flow computations on the 
graphs we expect to see in practice. However, a simple pre-processing pruning step, run on $H$ before augmenting it, typically results 
in massive reduction in the size of the graph on which the max-flow computation must be run. 

Consider node $a_w$ in the example graph in Figure \ref{fig:examplepushpull}(ii). 
The best decision for that node by itself is a {\em push} decision (since $PUSH(a_w)$ $=$ $3 < PULL(a_w) = 10$). Since there is no 
node upstream of $a_w$ (which is a writer node), we can assign this node a {\em push} decision without affecting decisions at any other
node (any node downstream of $a_w$ can still be assigned either decision), and remove it from the graph. Similarly we can assign 
push decision to node $b_w$ and remove it from $H$. After that, we can see that node
$i_1$ can also now be assigned a {\em push} decision (optimal for it in isolation) without 
affecting any other node. Similarly, we can assign {\em pull} decisions to nodes $m_r, n_r, p_r$, $q_r$ and remove them by an analogous reasoning. 



We now state the pruning rules, which are applied directly to $H$ (i.e., before constructing the augmented graph): \\[2pt]
                                                                                                                      (P1) recursively remove all 
nodes $v$ such that $w(v) > 0$ and $v$ has no incoming edges, and assign them {\em push} decisions, \\[2pt] (P2) recursively remove all nodes $v$ such that $w(v) < 0$ and $v$ has no
outgoing edges, and assign them {\em pull} decisions.\\[2pt]  
This pruning step can be applied in linear time over the overlay graph. 
We show in Section~\ref{sec:dataflowdecision} that the graphs after pruning is much smaller
than the original graph, and is usually disconnected with a large number of small connected components. We apply the above max-flow-based algorithm to each of 
the connected components separately.
\vskip 2pt

\begin{theorem}
Use of pruning rules P1 and P2 does not compromise optimality.
\end{theorem}
\vskip 8pt
{\bf Proof:} Let $H$ denote the original overlay graph, and let $H^p$ denote the pruned graph after applying rules P1 and P2. 
Let $H'$ denote the augmented graph constructed from $H$. Let $v$ denote a node that was pruned using rule P1, and thus 
assigned a {\em push} decision. Let $UPSTREAM(v)$ denote the set of all nodes upstream of $v$ in $H$ (including $v$). 
According to P1, node $v$ being pruned and assigned a {\em push} decision implies that all nodes in $UPSTREAM(v)$ must also been pruned  and assigned a {\em push} decision.
It follows from definition of $UPSTREAM(v)$ that there are no directed edges into $UPSTREAM(v)$ from any of the remaining
nodes in $H$.
Further, for any node $u \in UPSTREAM(v)$, $w(v) > 0$ (for P1 to apply), 
and thus there is an edge $(u, t)$ in $H'$ but no edge from $s$ to $u$. Thus, even in $H'$, there is no edge 
to a node in $UPSTREAM(v)$ from one of the remaining nodes $H'$.

Putting these together, we have that the nodes in $UPSTREAM$ $(v)$ are not reachable from $s$ in $H'$. Thus in any network
flow that originates at $s$, no flow can reach the nodes in $UPSTREAM(v)$ and those nodes do not participate in the network
flow. As a result, the nodes in $UPSTREAM(v)$ will remain unreachable from $s$ in the residual graph at the end of max-flow algorithm, and 
hence, they will all be assigned
%
a {\em push} decision in the end (i.e., same decision as assigned by the pruning step). By an analogous reasoning,
we can show that the nodes that are pruned by the pruning rule P2 and assigned a {\em pull} decision will remain unreachable from 
$s$ in the residual graph of max-flow as well (this is because there is no path from those nodes to $t$, and hence they do not participate in the 
max-flow solution either).

Further, since the pruned nodes do not participate in the max-flow in any way, removing them does not change the max-flow solution 
found either. Hence, the decisions made on the pruned graph $H^p$ will be identical to the ones made on the original graph $H$.

\noindent
$\square$


\label{algo:split}

\subsection{Greedy Alternative to the Max-flow-based Algorithm} 
Although we found the pruning step to be highly effective in reducing the complexity of the
max-flow-based algorithm, we also sketch a simpler greedy algorithm for making dataflow decisions in
case the pruning step results in a very large connected component (we note that we did not encounter
such a scenario in our extensive experimental evaluation). We traverse the overlay graph starting
from the writers in a breadth-first manner. After processing a node, it may be assigned one of
three decisions: (1) {\em push}, (2) {\em pull}, (3) {\em tentative pull}. 
A {\em tentative pull} decision may be
changed to a {\em pull} or {\em push} decision, but a {\em pull} or a {\em push} decision, once
made, is final.
We maintain two invariants at all times: (1) a node that is assigned a {\em tentative pull} decision is never downstream of a node assigned 
a {\em pull} or {\em tentative pull} decision, (2) a node assigned a {\em push} decision is never
downstream of a node assigned a {\em pull} or a {\em tentative pull} decision.
When processing a node $v$:
\begin{itemize}
\item If one of the input nodes to $v$ has been assigned a {\em pull} decision, we assign a {\em pull}
decision to $v$.
\item If $PUSH(v) > PULL(v)$ (i.e., the node should be assigned a {\em pull} decision) and at least one of
$v$'s input nodes is assigned a {\em tentative pull} decision, then we assign a {\em pull}
decision to $v$. We also change the {\em tentative pull} decisions assigned to any of its input
nodes to {\em pull}.
\item If $PUSH(v) > PULL(v)$ (i.e., the node should be assigned a {\em pull} decision) and none of
its input nodes are assigned {\em pull} or {\em tentative pull} decisions, we assign a {\em
    tentative pull} decision to $v$.
\item If $PUSH(v) < PULL(v)$ (i.e., the node should be assigned a {\em push} decision) and all of its
input nodes are assigned a {\em push} decision: we assign a {\em push} decision to $v$.
\item Finally, if $PUSH(v) < PULL(v)$ and some of its input nodes are assigned a {\em tentative pull} decision, then 
we make a greedy local decision for all of those nodes together. That is, we check the total cost of
assigning all of those nodes a {\em push} decision and the total cost of assigning all of them a
{\em pull} decision, and pick the best among the two. 
\end{itemize}

It is easy to see that the algorithm maintain the two invariants at all times, and thus produces a 
valid set of dataflow decisions at the end (at the end, {\em tentative pull} decisions, if any, are
treated as {\em pull} decisions). This greedy algorithm runs in time linear in the number of edges (each edge is processed at most
twice), and is thus highly efficient. 

%

\subsection{Partial Precomputations by Splitting Nodes}
\begin{figure}[t]
\centering
 \includegraphics[width=92mm]{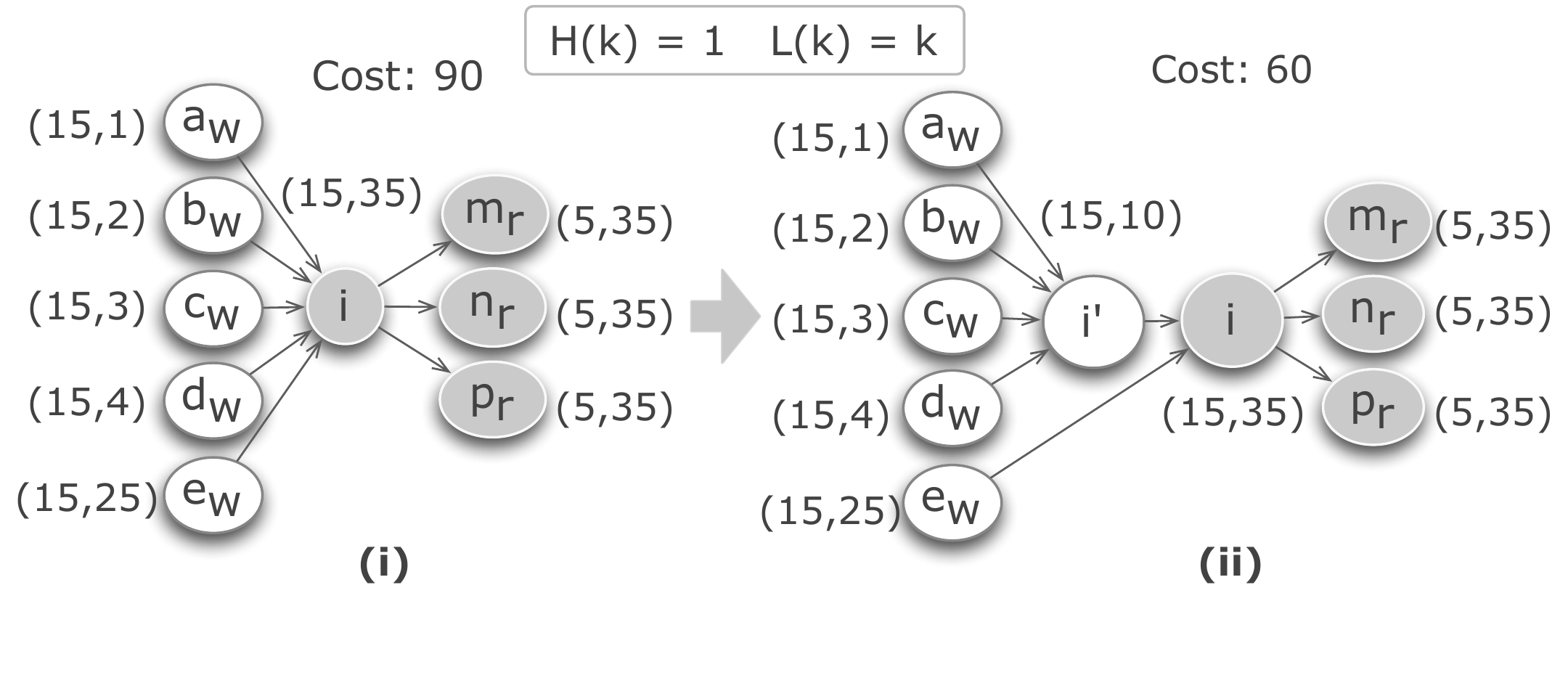}
 \vspace{-20pt}
 \caption{Splitting a node based on push-pull frequencies.}
\label{fig:splitpushpull}
\end{figure}

Making decisions on a per-node basis can lose out on a significant optimization opportunity -- based on the 
push and pull frequencies, it may be beneficial to partially aggregate a subset of the inputs to an aggregate
node. Figure~\ref{fig:splitpushpull} shows an example. Here, because of the low write frequencies for inputs $a_w, b_w, c_w,$ and $d_w$ for aggregator node $i$, it is better
to compute a partial aggregate over them, but compute the full aggregate (including $e_w$) only when needed (i.e., on a read).

One option would be to make the pre-computation decisions on a per-edge basis. However, that optimization 
problem is much more challenging because the cost of an incremental update for an aggregate node depends on how many
of the inputs are being incrementally aggregated, and how many on demand; thus the decisions made for different edges are not
independent of each other. Next we propose an algorithm that achieves the same goal, but in a much
more scalable manner. 

For every node $v$ in the overlay graph, we consider splitting its inputs into two groups. Let $f$
denote the pull frequency for $v$, and let $f_1,
..., f_k$ denote the push frequencies of its input nodes, sorted in the increasing order. For every
prefix $f_1, ..., f_l$, of this sequence, we compute: $\sum_{i \le l} f_i H(l) + f \times L(l)$. 
We find the value of $l$ that minimizes this cost; if $l \ne 0$ and $l \ne k$, we construct a new
node $v'$ that aggregates the inputs corresponding to frequencies $f_1, ..., f_l$, remove all those
inputs from $v$, add $v'$ as an input to $v$. As we show in our experimental evaluation, this
optimization results in significant savings in practice.

\subsection{Adapting the Dataflow Decisions}
Most real-world data streams, including graph data streams, show significant variations in read/write 
frequencies over time. We propose and empirically evaluate, a simple adaptive scheme to handle 
such variations. For a subset of the overlay nodes (specified below), we monitor the observed push/pull 
frequencies over recent past (the window size being a system parameter).
If the observed push/pull frequencies at a node are significantly different than the estimated frequencies,
then we reconsider the dataflow decision just for that node and change it if deemed beneficial. 
Dataflow decisions can be unilaterally changed in such a manner only for: {\em pull} nodes all of whose
upstream nodes are designated {\em push}, and {\em push} nodes all of whose downstream nodes are designated {\em pull}
(we call this the {\em push/pull frontier}).
Hence, these are the only nodes for which we monitor push/pull frequencies (it is also easier to maintain
the push/pull frequencies at these nodes compared to other nodes). Techniques for more sophisticated adaptive schemes is a rich area which we plan to pursue in future.


\section{Evaluation}
\label{sec:evaluations}

In this section, we present a comprehensive experimental evaluation using our prototype system using
several real-world information networks. Our results show that overall our approach results in
significant improvements, in many cases order of magnitude improvements, in the end-to-end throughputs
overall baselines, and that our overlay construction algorithms are effective at finding compact
overlays. 

\begin{figure*}[th]
\centering
\includegraphics[width=180mm]{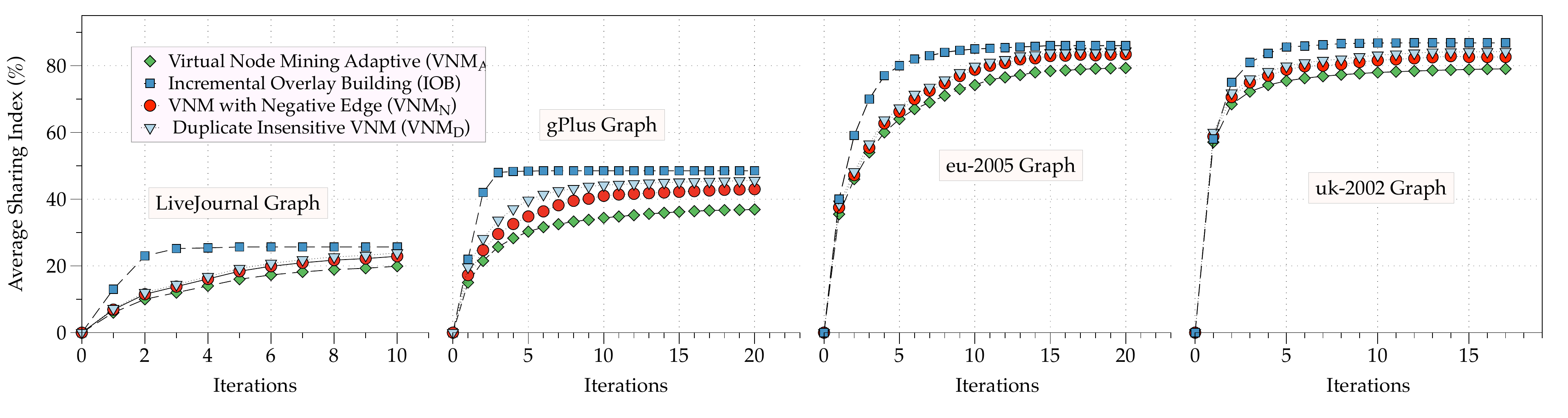}
\vspace{-4pt}
\caption{ Comparing overlay construction algorithms on real networks; (\iob{} should not be directly compared against \vnmn{} or \vnmd{} since it does not use negative edges or duplicate paths)}
\label{fig:compare-runtime}
\vspace{-4pt}
\end{figure*}

\subsection{Experimental Setup}
We ran our experiments on a 2.2GHz, 24-core Intel Xeon server with 64GB of memory, running 64-bit
Linux. Our prototype system is implemented in Java. We use a set of  dedicated threads to play back 
the write and read traces (i.e., to send updates and queries to the system), and a thread pool to 
serve the read and write queries. 


\topicu{Datasets and Query Workload}
We evaluated our approach on several real-world information networks including~\footnote{\scriptsize First three
are available at \url{http://snap.stanford.edu/data/index.html}, and the latter three at
\url{http://law.di.unimi.it/}.}: \\[2pt]
(1) LiveJournal social network ({\em soc-LiveJournal1}: 4.8M nodes/69M edges), \\
    (2) social circles from Google+ ({\em ego-Gplus}:107k/13M), \\
(3) Web graph of Berkeley and Stanford ({\em web-BerkStan}: 685k/7.6M), \\
(4) {\em Hollywood-2009} Social Graph (1.1M/114M), \\
(5) {\em EU2005} Web Graph (862k /19M),
and  \\
(6) {\em UK2002} Web Graph (18.5M /298M). \\ 
In this paper, we report results for the
first two, and the last two. 

We report results for three ego-centric aggregate queries: \suma{}, \maxa{}, and \topka{}, all
specified over 1-hop neighborhoods. \suma{} and \maxa{} queries ask us to compute the total sum and 
the max over the input values respectively. \topka{} asks for the $k$ most {\em frequent}
values among the input values, and is a holistic aggregate~\cite{tag}.\footnote{\scriptsize In other words, \topka{} is 
a generalization of {\em mode}, not {\em max}.}

Since the user activity patterns (i.e., read/write frequencies) are not
available for any real-world network that we are aware of, we generate those synthetically using a
Zipfian distribution; event rates in many applications like tweets
in Twitter, page views in Yahoo!'s social platform have been shown to follow a Zipfian distribution~\cite{feedingfrenzy,zipfweb}.
Further, we assume that the read frequency of a node is linearly related to its write frequency; we
vary the write-to-read ratio itself to understand its impact on the overall performance.
For some of the experiments, we used real network packet traces to simulate user activity~\footnote{\scriptsize Available at \url{http://ita.ee.lbl.gov/html/traces.html}.}: (1) EPA-HTTP, and (2) UCB Home IP Web Traces.

\topicu{Evaluation Metric}
Our main evaluation metric is the {\em end-to-end throughput of the system}, i.e., the total number of read
and write queries served per second. 
This metric accounts for the side effects of all potentially unknown system parameters whose impact might not show 
up for a specifically designed metric, and thereby reveals the overall efficacy of the system. 
When comparing the overlay construction algorithms, we also use the following metrics: sharing
index (SI), 
memory consumption, and running time. 

\eat{As our {\em ego-centric aggregation query execution} is composed of two different objectives, i.e., {\em construction of the overlay} and {\em execution of the queries on the overlay}, we carry our evaluations in the same flow as well. At first,  we evaluate the algorithms that we have proposed to build the overlay structure from the data graph. Then we evaluate the efficiency of query execution. For the first part, we start with a set of synthetic graphs, as it's easier to vary parameters (in our case it's the preferential attachment factor) of the graph generator in this case.}


\topicu{Comparison Systems or Algorithms} For overlay construction, we compare five algorithms:
\vnm{}, \vnma{}, \vnmn{}, \vnmd{}, and \iob{}.
For overall throughput comparison, we compare three approaches: 
(1) {\em all-pull}, where all queries are evaluated on demand (i.e., no sharing of aggregates and no
pre-computation), (2) {\em all-push}, where all aggregates are pre-computed, but there is no sharing
of partial aggregates,  and (3) {\em dataflow-based overlay}, i.e., our approach with sharing of aggregates and selective pre-computation. 
We chose the baselines based on industry standards: {\em all pull} is typically seen in social
networks,
whereas {\em all push} is more prevalent in data streams and
complex event processing (CEP) systems.

\vspace{8pt}
\subsection{Overlay Construction}
\topicu{Sharing Index} First we compare the overlay construction algorithms with respect to the {\em
average sharing index} achieved per iteration, over 5 runs (Figure \ref{fig:compare-runtime}). 
As we can see, \iob{} finds more compact overlays (we observed this consistently for all graphs that
we tried). The key reason is that: \iob{} considers the entire graph when looking for sharing
opportunities, whereas the \vnm{} variations consider small groups of readers and writers based on
heuristical ordering of readers and writers. 
Note that, \iob{} should only be compared against
\vnma{}, and not \vnmn{} or \vnmd{}, since it does not use negative edges or duplicate paths.
We also note that, for \iob{}, most of the benefit is obtained in first few iterations,
whereas the VNM-based algorithms require many iterations before converging.
Further, the overlays found by \vnmn{} and \vnmd{} are significantly better than those
found by \vnma{}.
This validates our hypothesis that using negative edges and reusing mined edges, if possible, 
     results in better overlays.
%
%
%
Another important trend that we see here is that the sharing indexes for web graphs are typically much
higher those for the social graphs. 
Kumar et al. also notice similar difficulties in achieving good structural compression in
social networks~\cite{shinglesravi}.

\begin{figure}[h]
\centering
\includegraphics[width=80mm]{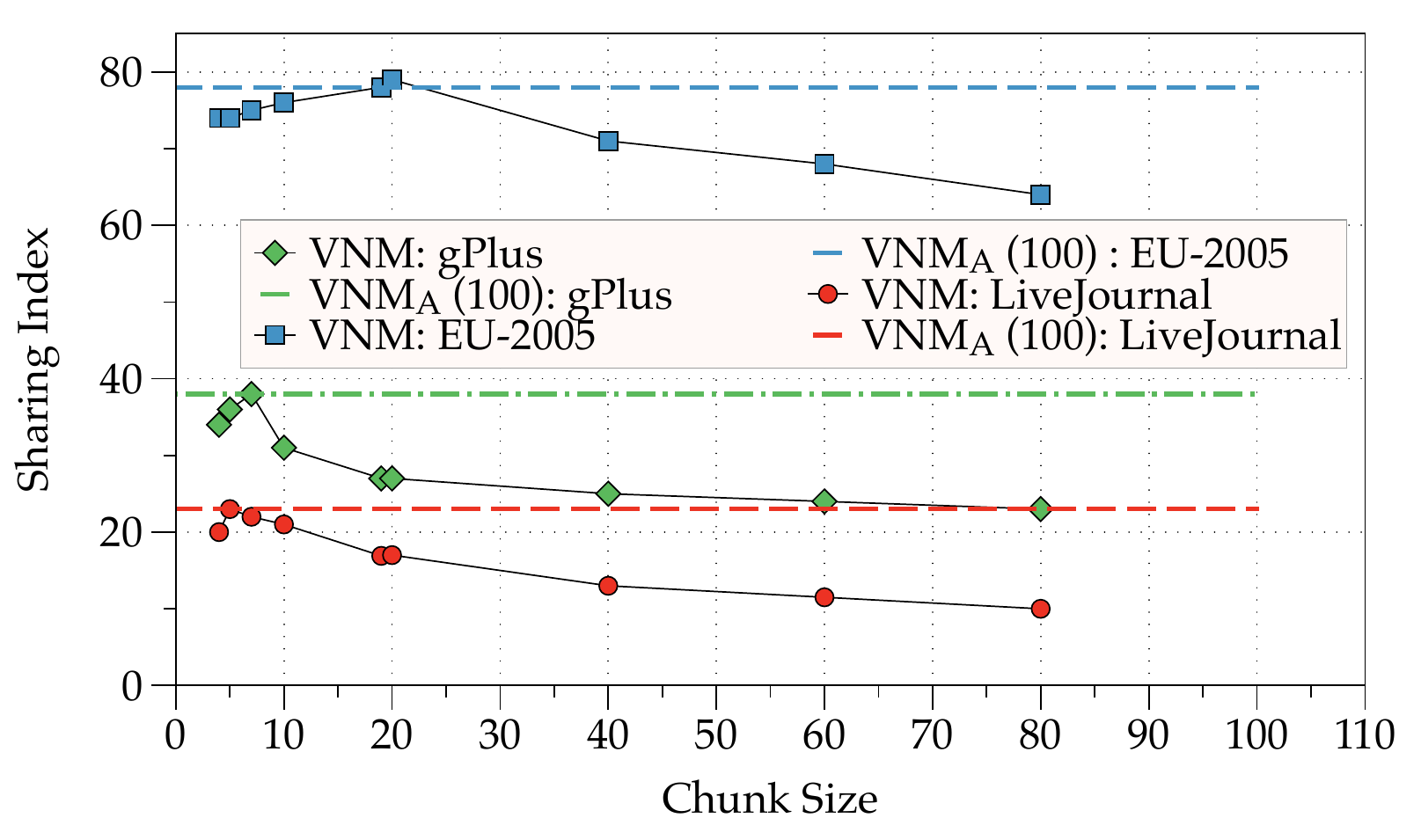}
\vspace{-8pt}
\caption{
Effect of chunk size on \vnm{} 
}
\label{fig:chunk}
\vspace{-8pt}
\end{figure}

\vspace{5pt}
\topicu{Comparing \vnm{} and \vnma{}} Figure \ref{fig:chunk} shows SI achieved by our adaptive \vnma{} algorithm and
by \vnma{}
as the chunk size is varied. As we can see, \vnm{} is highly sensitive to this
parameter, whose optimal value is quite different for different data graphs. On the other hand, \vnma{} is able
to achieve as compact an overlay (in some cases, slightly better) as the best obtained by \vnm{}.

\begin{figure}[th]
\centering
  \includegraphics[height=42.5mm]{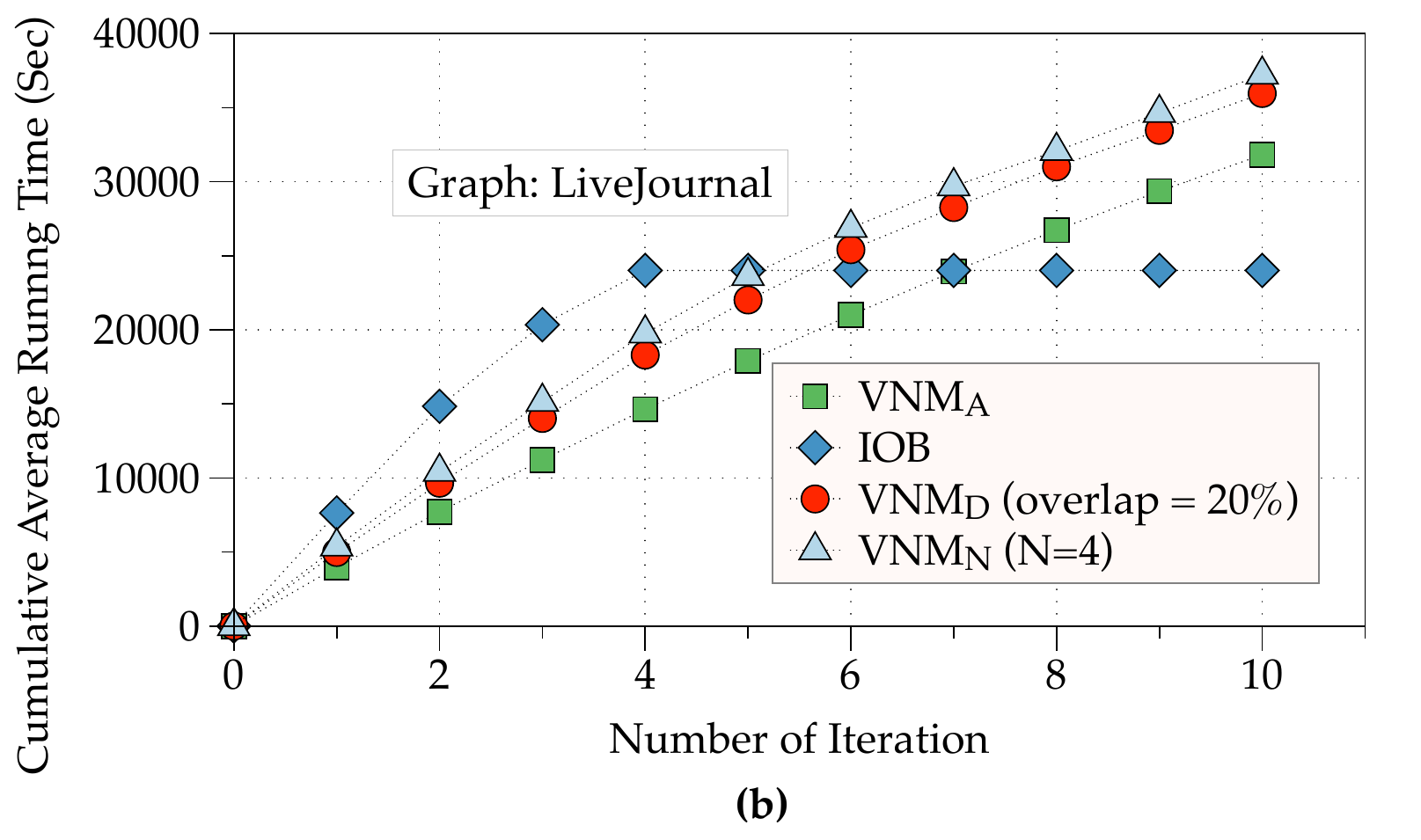}
 \includegraphics[height=44.5mm]{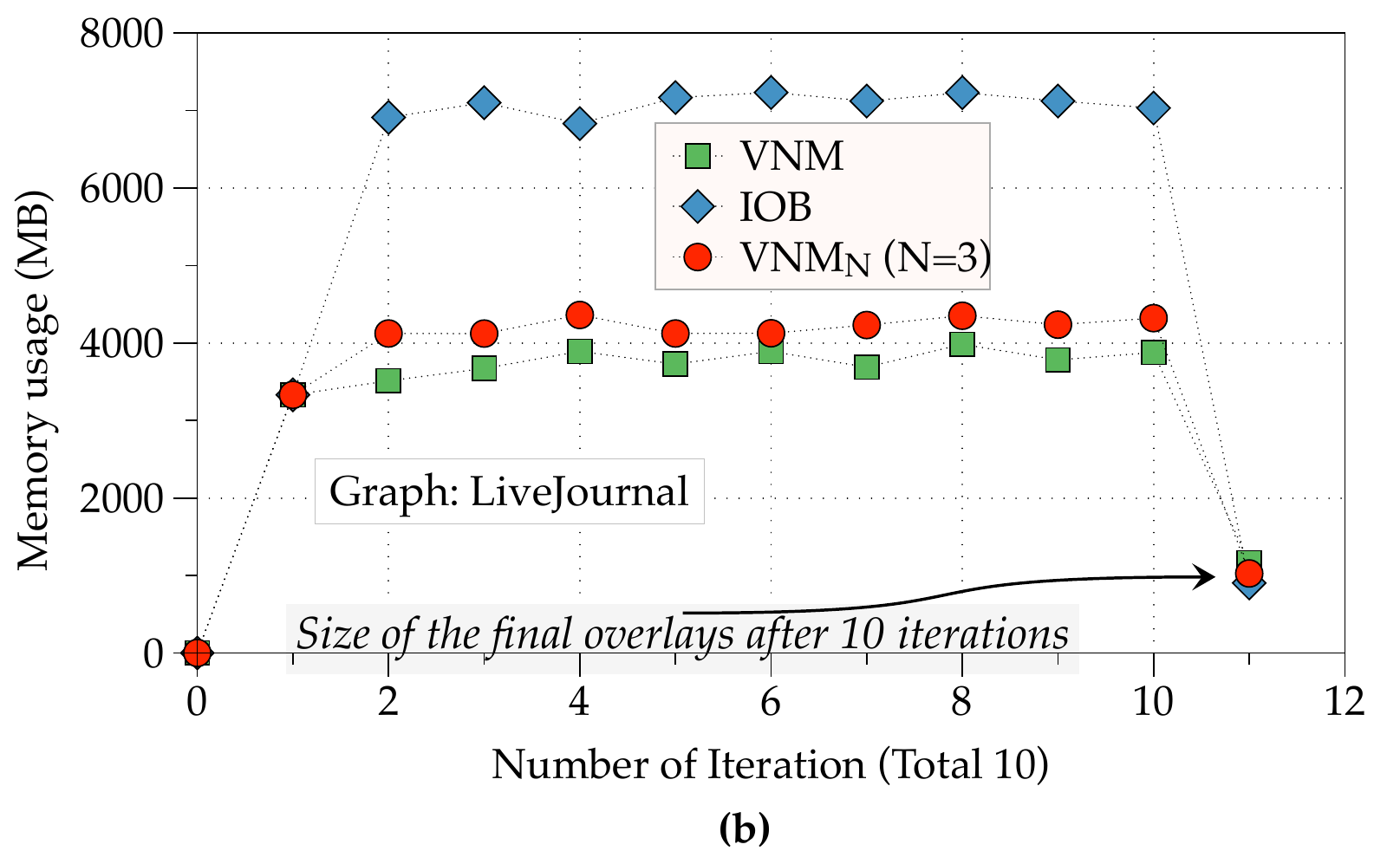}
 \vspace{-8pt}
\caption{ (a) Running time comparison of the overlay construction algorithms; (b) Memory consumption comparison of the overlay construction algorithms.
}
\vspace{-8pt}
\label{fig:runtime-memory}
\end{figure}

\topicu{Running Time and Memory Consumption} Figure~\ref{fig:runtime-memory}(a) shows the running time for the different
construction algorithms with the increasing number of iterations for the LiveJournal graph.
As we can see \iob{} takes more time for first few iterations, but is overall faster than the
\vnma{} and its variations since it converges faster. As expected, both \vnmn{} and \vnmd{} take
more time per iteration than \vnma{}.
We also compared ( Figure~\ref{fig:runtime-memory}(b)) the total memory consumption of the overlay construction algorithms . For
LiveJournal, \vnma{} and its variations used approximated 4GB of memory, whereas \iob{} used 8GB at
its peak; this is not surprising considering that \iob{} needs to maintain additional global
data structures.

\begin{figure}[h]
\centering
\includegraphics[height=41.5mm]{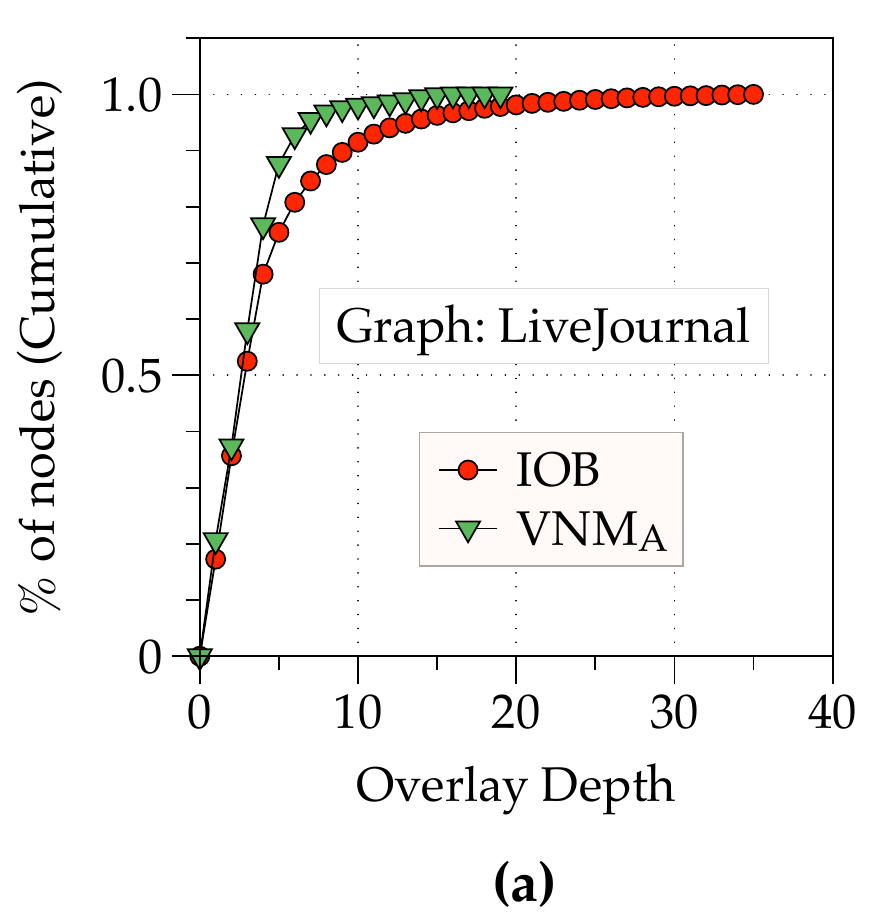}
\includegraphics[height=41.5mm]{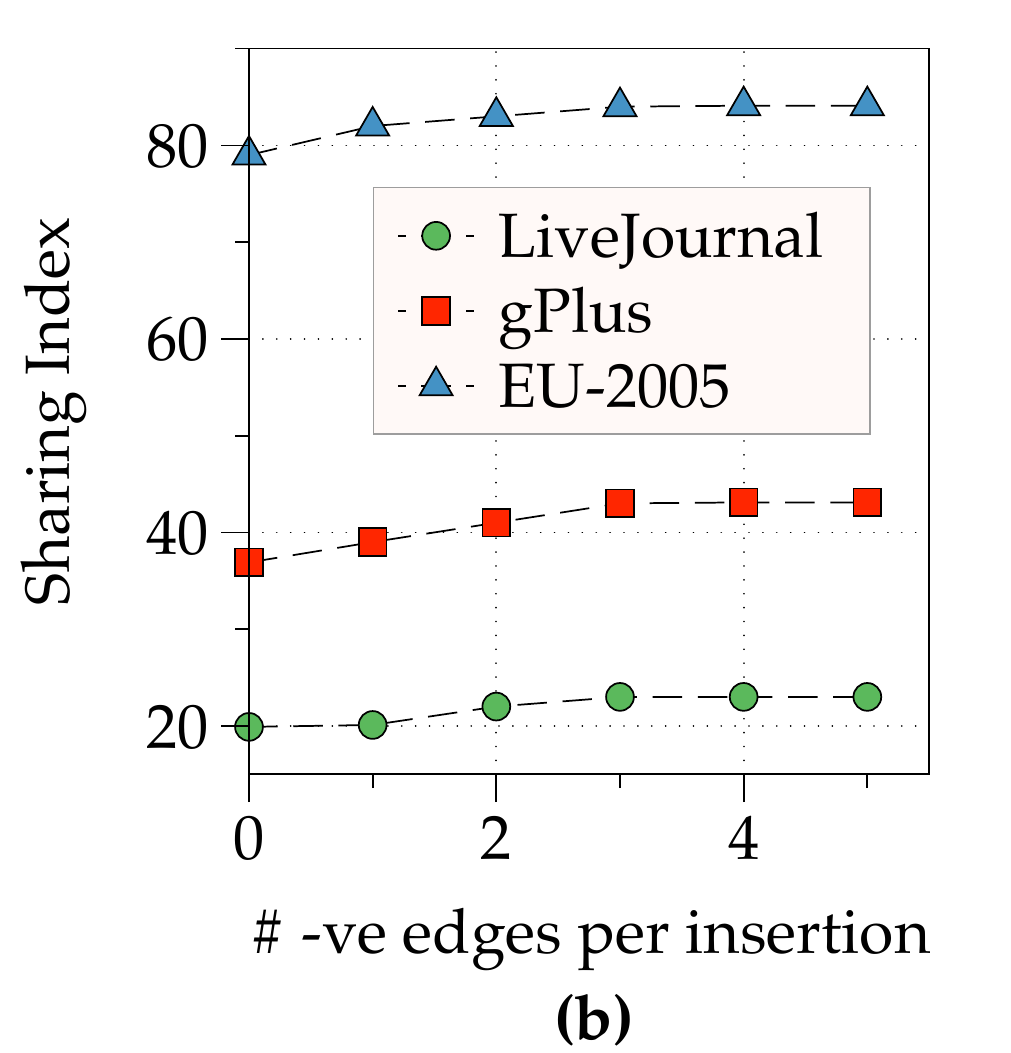}
\vspace{-8pt}
\caption{  (a) Comparison of overlay depth for different overlay construction algorithms. (b) Effect of increasing parallelism on throughput.}
\label{fig:overlay-depth}
\vspace{-8pt}
\end{figure}

\topicu{Overlay Depth} Figure~\ref{fig:overlay-depth}(a) compares the depths of the overlays created by
\vnma{} and \iob{} algorithms for one run over the LiveJournal graph. The overlay depth for a reader
is defined to be the length of the longest path from one of its input writers to the reader. In
the figure, we plot the cumulative distribution of the number of readers at each overlay depth. As
we can see, \iob{} creates a significantly deeper overlay with average depth of 4.66 (vs 3.44 for \vnma{}); as we will see later, this results in
lower end-to-end throughput, even though \iob{} creates a more compact overlay.


\begin{figure}[t]
\centering

 \includegraphics[height=42.5mm]{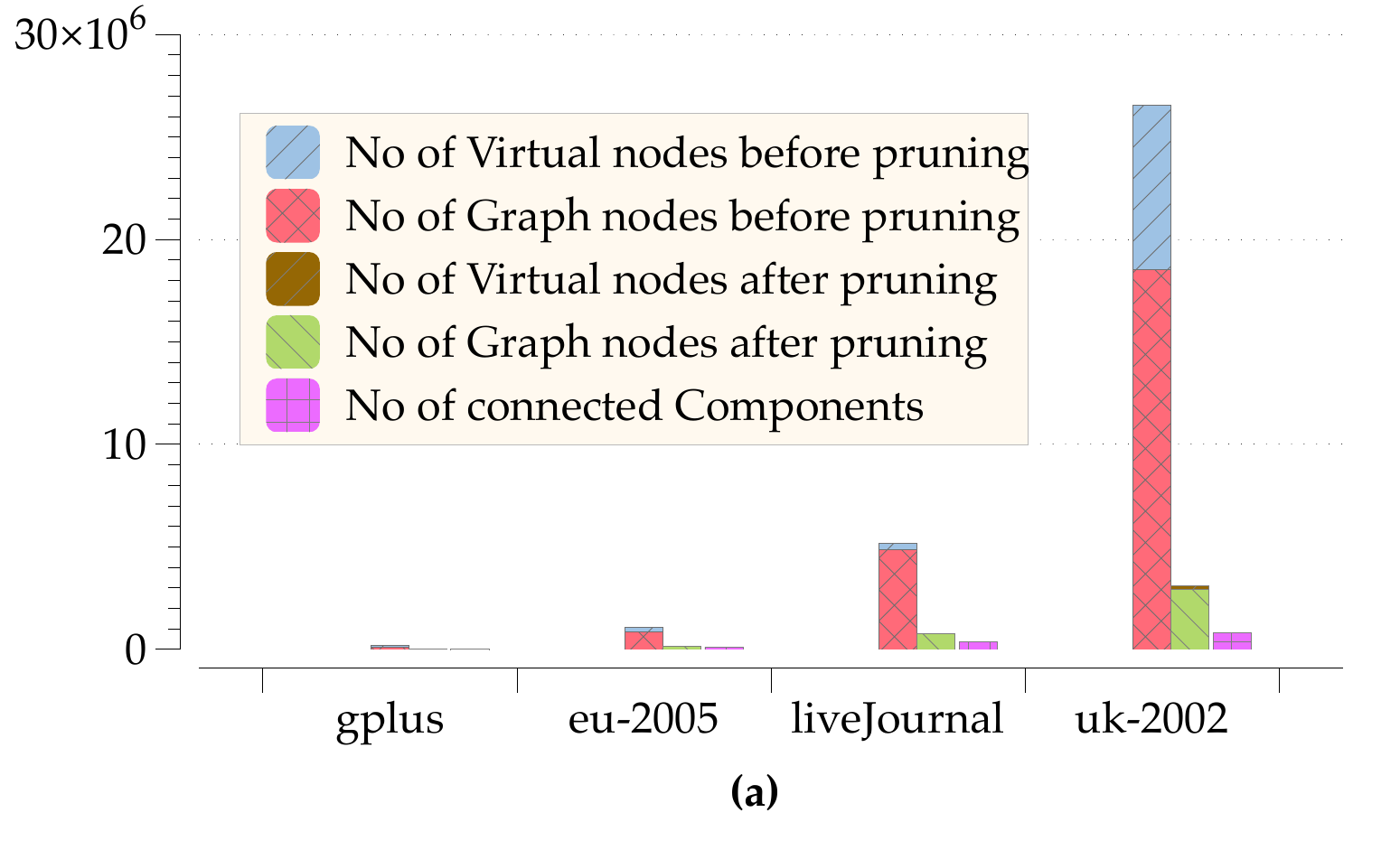}
  \includegraphics[height=44.5mm]{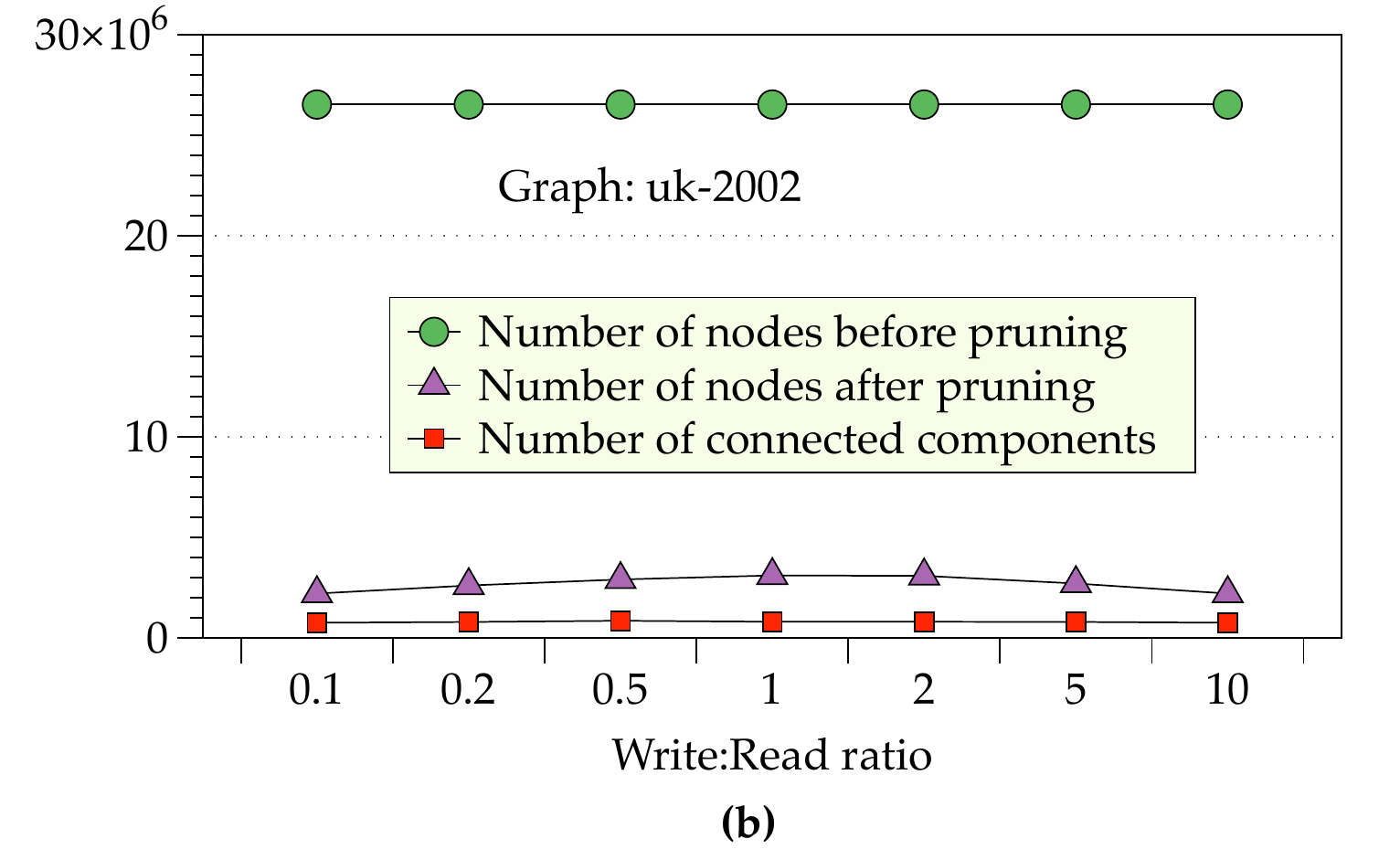}
\vspace{-8pt}
\caption{(a) Benefits of pruning before running maxflow for various graphs with 1:1 write-to-read ratio;
(b) Benefits of pruning before running maxflow for different write-to-read ratios on uk-2002 graph.
}
\label{fig:pruning}
\vspace{-8pt}
\end{figure}

\topicu{Microbenchmarking \vnmn{}} Next we examine the \vnmn{} algorithm and the impact of the number
of negative edges allowed per transaction on the sharing index and running time. We varied the
number of allowed negative edges from 1 to 5. As we can see in Figure~\ref{fig:overlay-depth}(b), 
allowing for negative edges has a significant impact on the sharing index, and as expected, we do
not see much benefit beyond 3 or 4 negative edges. The running time of \vnmn{} increases rapidly as
we increase the number of negative edges allowed, and almost doubles when we allow 5 negative edges
vs none (not plotted). However, we note that overlay construction is a one-time process, and the
benefits in terms of increased throughput outweigh the higher initial overlay construction cost.

\begin{figure*}[t]
\centering
\includegraphics[height=49mm]{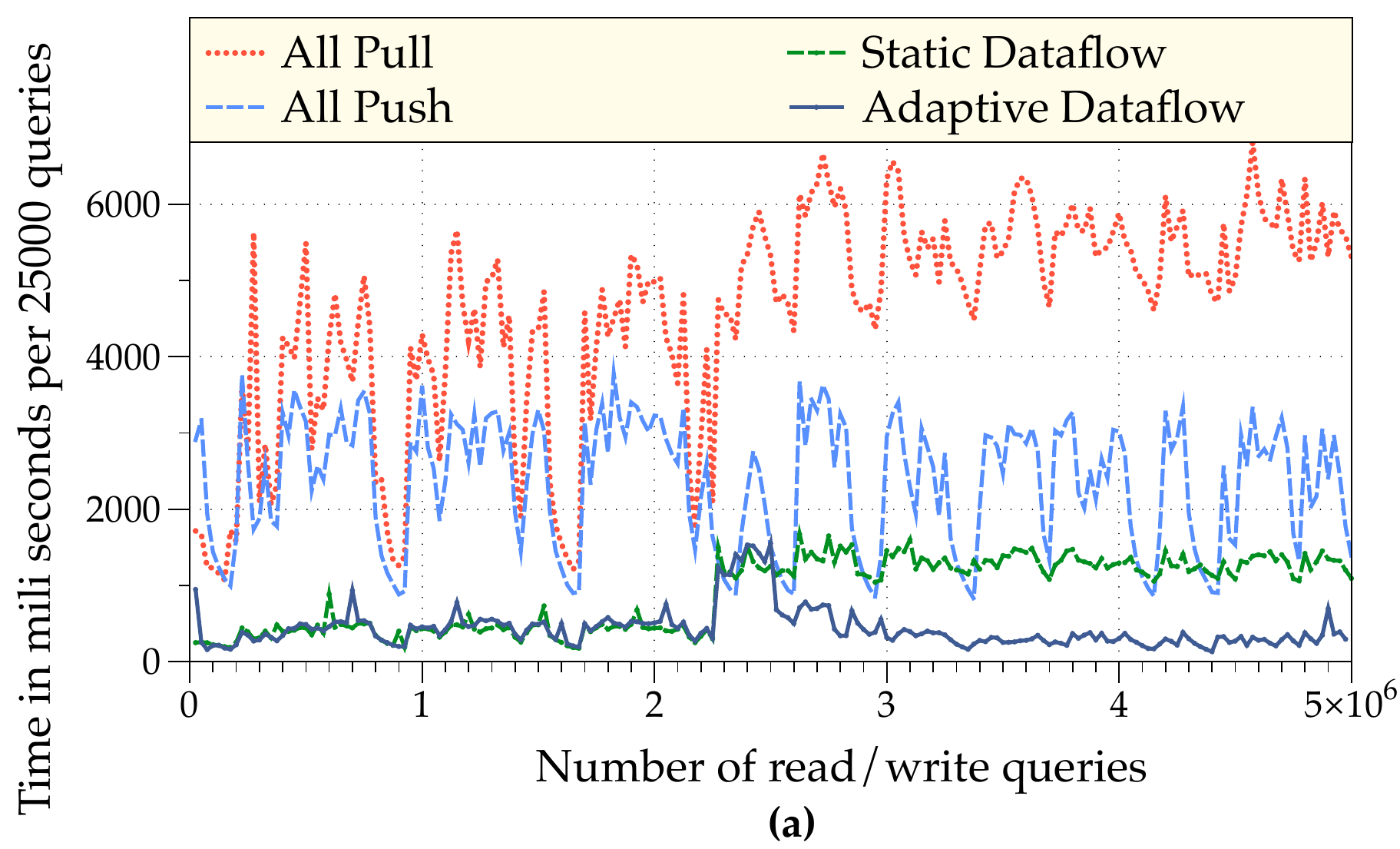}
\hspace{27pt}
\includegraphics[height=49mm]{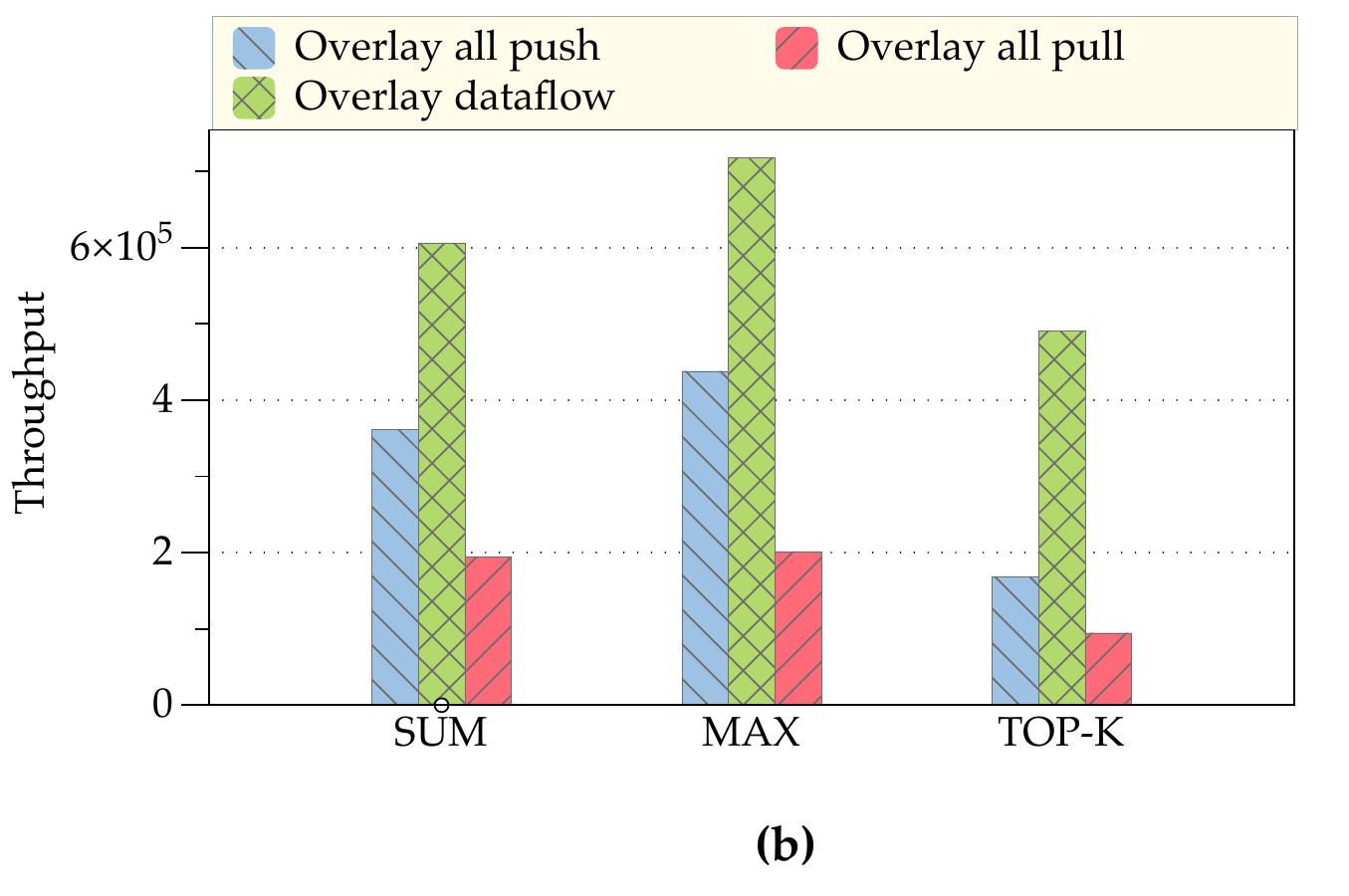}

\vspace{5pt}
\includegraphics[height=49mm]{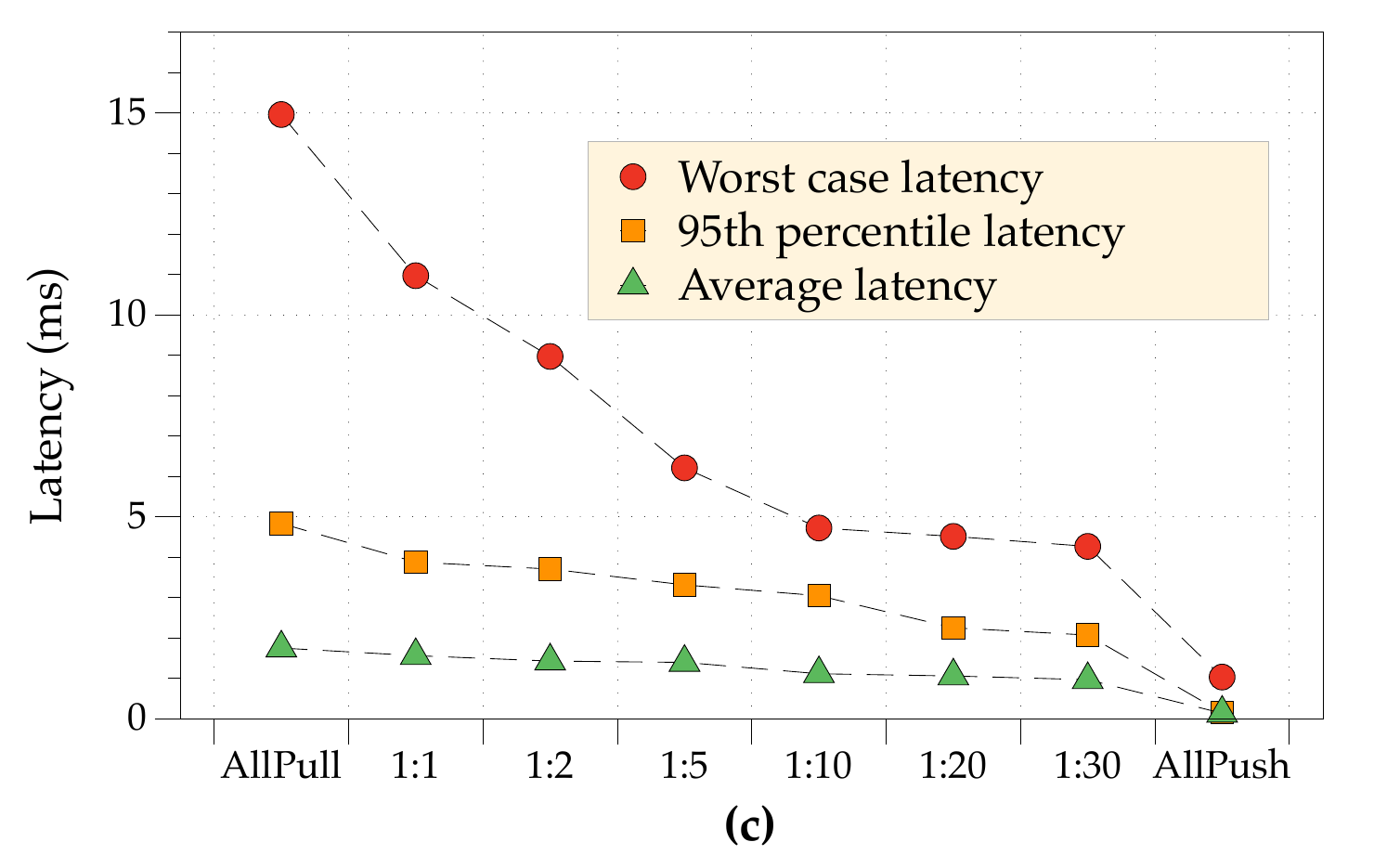}
\hspace{30pt}
\includegraphics[height=49mm]{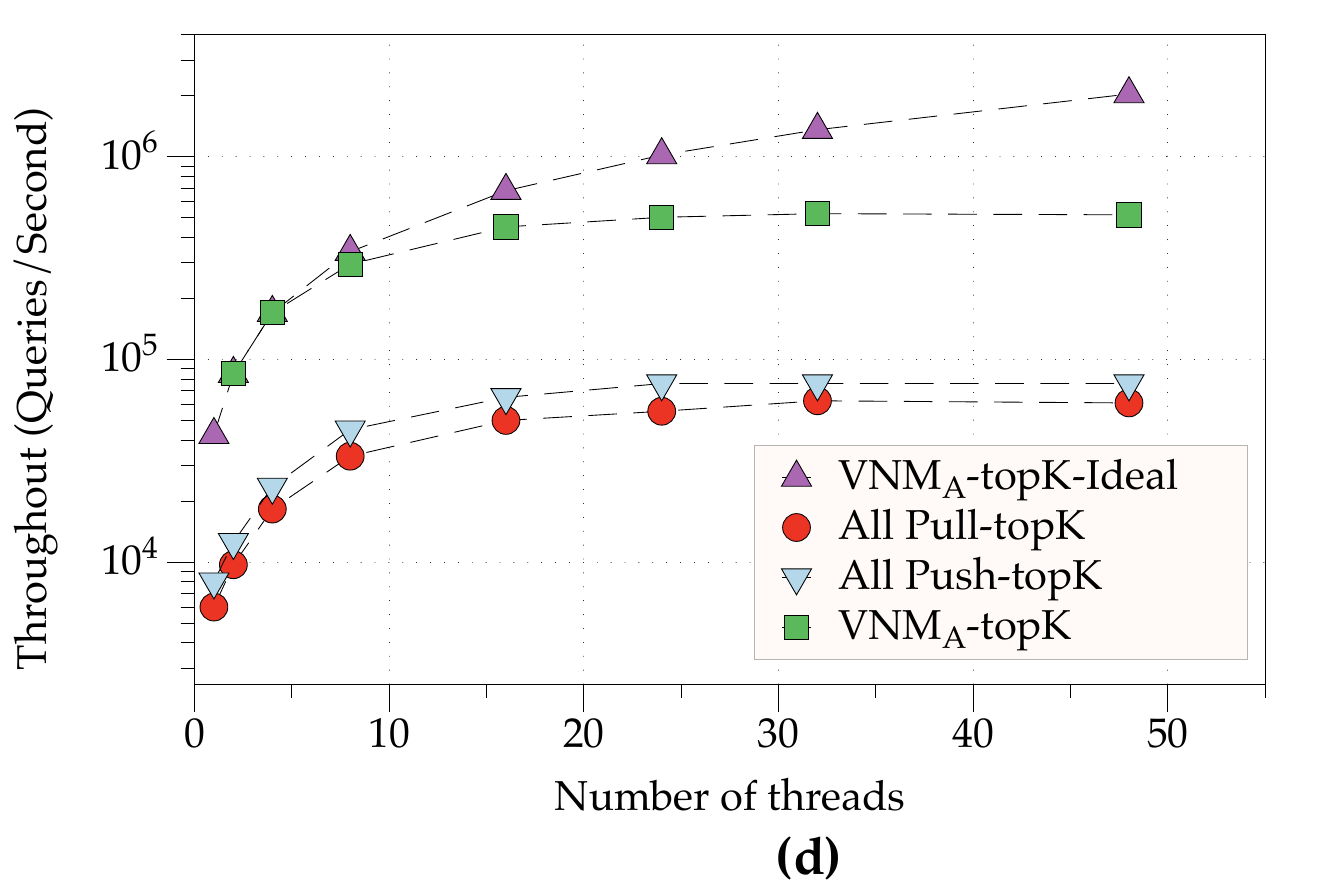}
\vspace{-8pt}
\caption{
(a) Effect of workload variations on different approaches;
(b) Baseline to motivate dataflow decisions;
(c) Read latency for different push:pull cost; 
(d) Effect of increasing parallelism on throughput.
} 
\label{fig:real-baseline-latency-scaleup}
\vspace{-8pt}
\end{figure*}

\begin{figure*}[t]
\centering
\includegraphics[height=50mm]{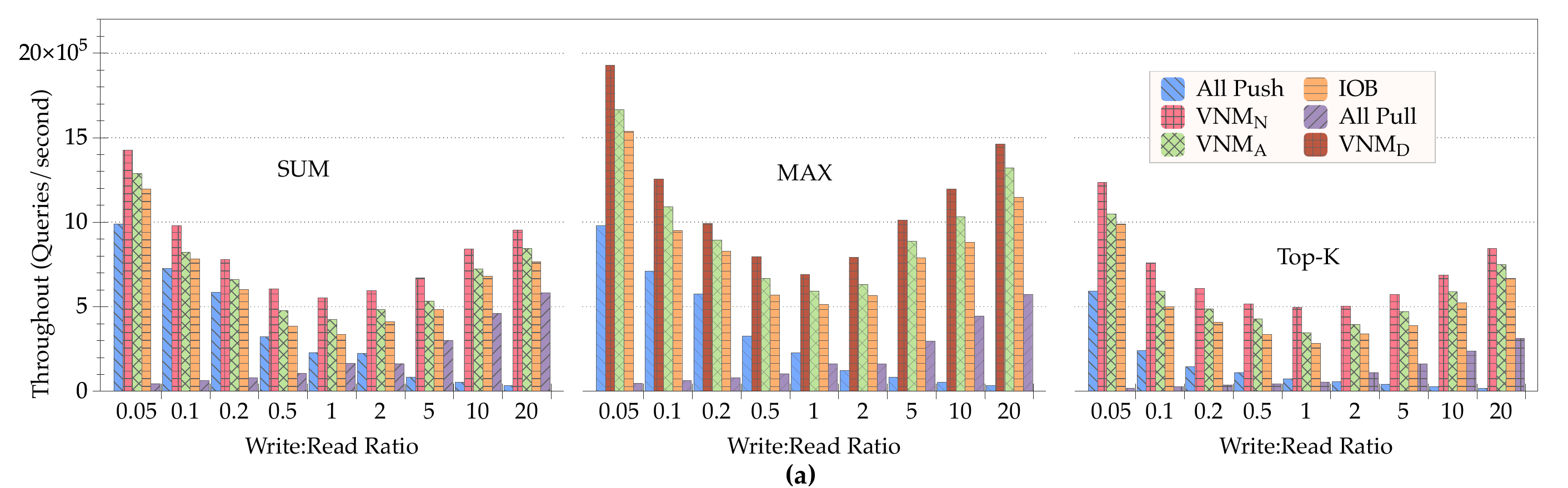}

\vspace{5pt}
\includegraphics[height=48.0mm]{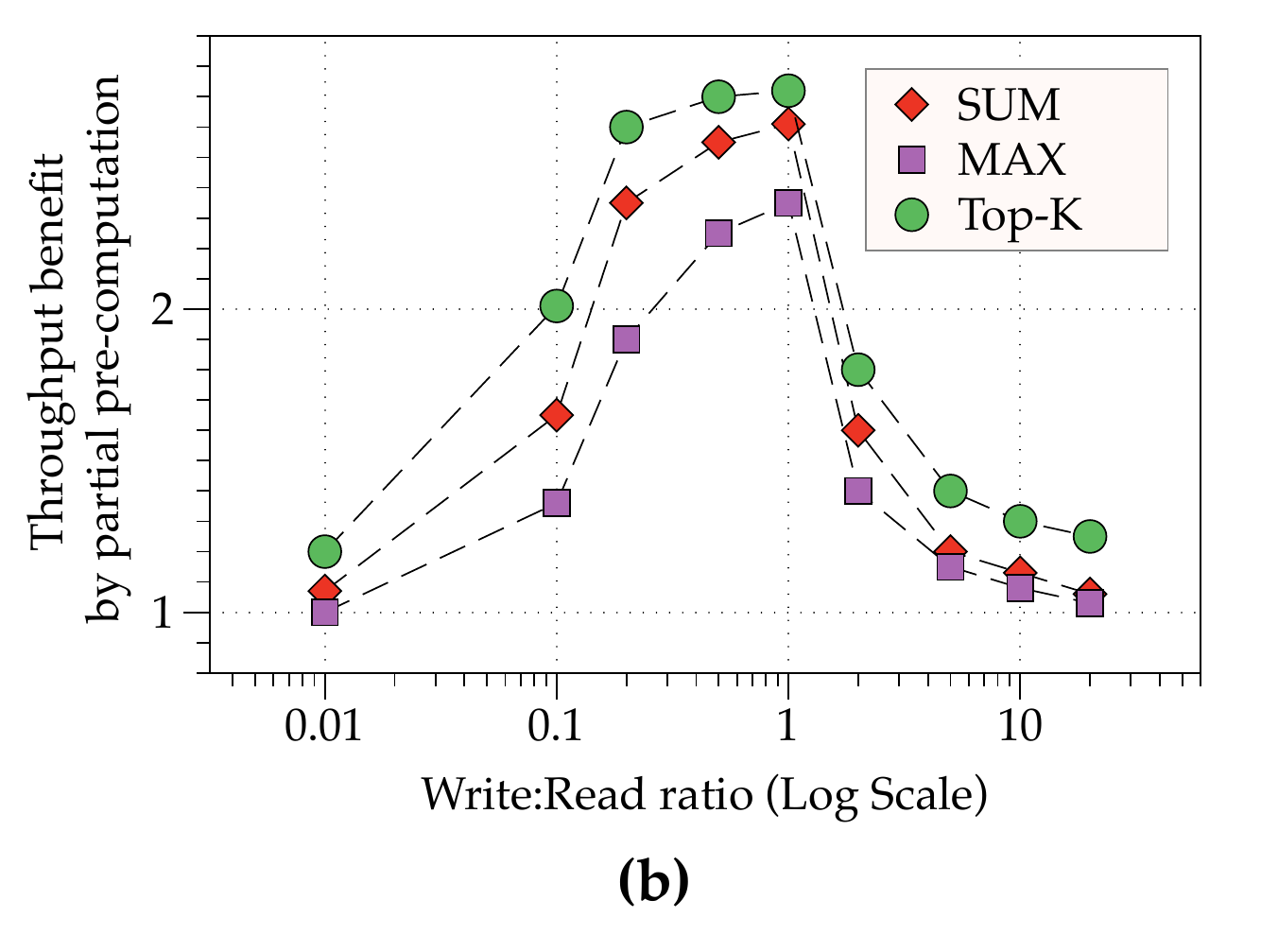}
\includegraphics[height=47.0mm]{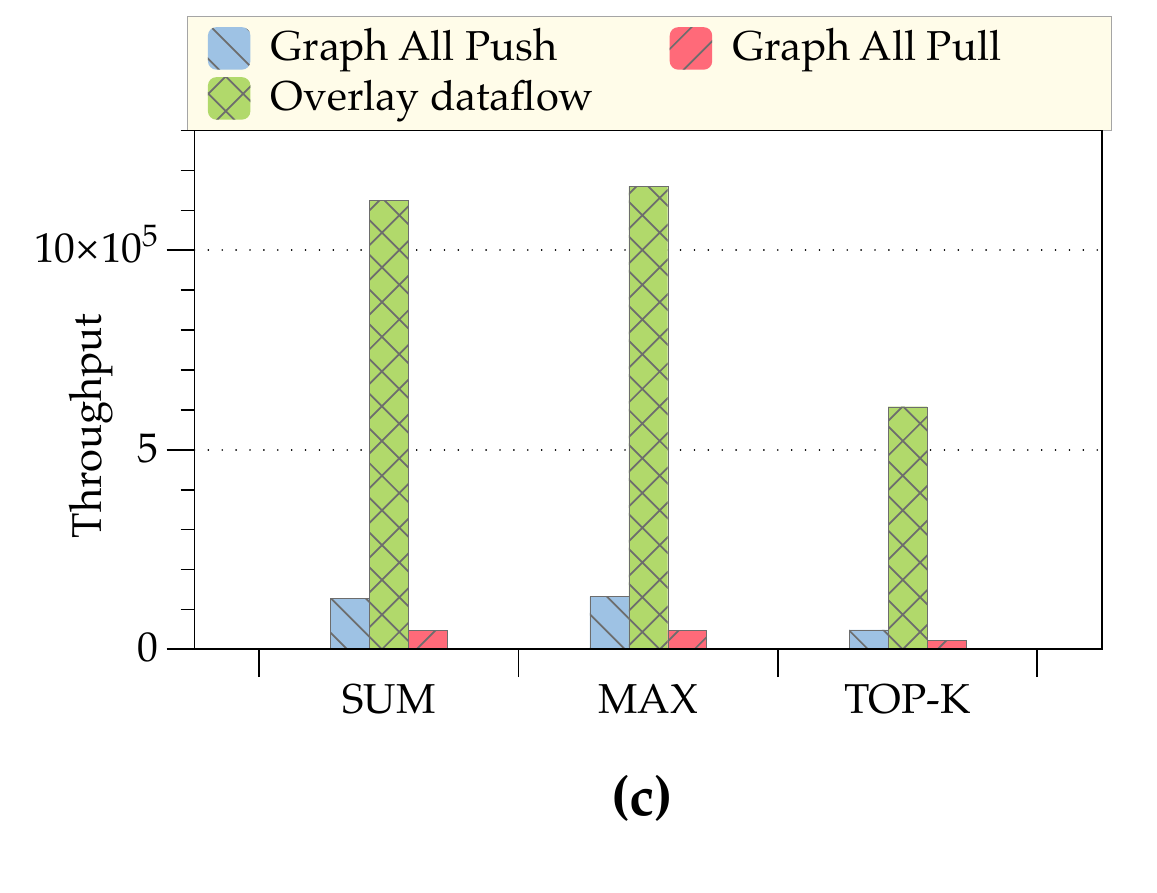}
\caption{
(a) End-to-end throughput comparison for different aggregate functions for the LiveJournal
graph, with 24 threads;
(b) Benefits of partial pre-computation through node splitting; 
(c) Throughput comparison for 2-hop aggregates.
}
\label{fig:throughput-split-2hop}
\end{figure*}

\subsection{Dataflow Decisions}
\label{sec:dataflowdecision}

\topicu{Effectiveness of Pruning} Figure~\ref{fig:pruning} shows the effectiveness of our 
pruning strategy to reduce the input to the max-flow algorithm. In the first part, we have plotted the result for a read/write ratio of 1:1 for different data graphs.
Each vertical bar in the figure has been divided to show the composition of intermediate overlay nodes and 
original graph nodes, before and after pruning the overlay that we got using \vnma{}.
We get similar results for other overlay construction algorithms. The pruning
step not only reduces the size of the graph (to below 14\% in all cases), but the resulting graph
is also highly disconnected with many small connected components, leading to very low running times for
the max-flow computations. In Figure~\ref{fig:pruning}(b), we show the same results but for different write:read
ratios for the uk-2002 data graph. As we can see, the benefits of pruning are lowest at
write-to-read ratio of 1, which is to be expected since the possibility of conflicts is highest
then. We saw similar results for the other graphs as well,
demonstrating the effectiveness of pruning as well as the scalability of max-flow-based dataflow decision making under various scenarios.

\topicu{Baseline for Dataflow Decisions} Figure~\ref{fig:real-baseline-latency-scaleup}(b) shows the effectiveness of the dataflow decisions on the overlay. In this experiment we kept the number of threads (12) and read/write ratio (1:1) of the queries fixed and computed the average throughput for: (a) overlay with all push, (b) overlay with dataflow decisions, and (c) overlay with all pull. As we can see, for all aggregate functions, overlay with optimal dataflow performs much better than overlay with all pull and all push thereby justifying our hypothesis. We observed similar results for other read/write ratios as well. 

\topicu{Adaptive Dataflow Decisions on a Real Trace} Figure~\ref{fig:real-baseline-latency-scaleup}(a) shows the ability of our proposed adaptive scheme 
to adapt to varying read/write frequencies.
We used the EPA-HTTP network packet trace to simulate read/write activity for nodes. We used average read/write frequencies of
the nodes to make static dataflow decisions. At a half-way point, we modified the read/write frequencies by increasing the read
frequencies of a set of nodes with the highest read latencies till that point. As we can see, the static dataflow decisions turn
out to be significantly suboptimal once this change is introduced. However, our simple adaptive approach is able to quickly adapt 
to the new set of read/write frequencies. 
%
%

\vspace{8pt}
\subsection{Throughput Comparison}

\topicu{Varying Read-Write Ratio}  Figure \ref{fig:throughput-split-2hop}(a) shows the results of our main
end-to-end throughput comparison experiments for the three ego-centric aggregate queries. We plot 
the throughputs for the two baselines as well
as for the overlays constructed by the different algorithms, as the write/read ratio changes from
0.05 (i.e., the workload contains mostly reads) to 20. As we can see, the overlay-based approaches consistently
outperform the baselines in all scenarios. For the more realistic write/read ratios (i.e., around
1), the throughput improvement over the best of the two baselines is about a factor of 5 or 6. For
read-heavy workloads, the overlay-based approach is multiple orders of magnitude better than the {\em
all-pull} approach, and about a factor of 2 better than the {\em all-push} approach, whereas the
reverse is true for the write-heavy workloads.

Comparing the different aggregate functions, we note that the performance improvements are much higher 
for the more computationally expensive \topka{} aggregate function. In some sense, simple aggregates like \suma{} and \maxa{} represent a worst
case for our approach; the total time spent in aggregate computation (which our approach aims to reduce through sharing) forms a smaller fraction of
the overall running time. 

Comparing the different overlay construction algorithms, we note that \vnmn{} shows significant
performance improvements over the rest of the overlay construction algorithms, whereas \iob{} is
typically the worst; the higher depth of the overlay increases the total amount of work that needs
to be done for both writes and reads.

\topicu{Effect of Splitting Aggregate Nodes} 
Figure \ref{fig:throughput-split-2hop}(b) shows the effect of our optimization of splitting an overlay aggregate
node based on the push frequencies of its inputs (Section ~\ref{algo:split}) on the LiveJournal graph.
As we can see, 
for all the aggregate functions, this optimization increases the throughput by more than a factor of 2 
when write-to-read ratio is around 1. In the two extreme cases (i.e., very low or very high
write-to-read ratios) where the decisions are either all push or all pull, this optimization has less
impact.

\topicu{Two-hop Aggregates}
Figure \ref{fig:throughput-split-2hop}(c) shows the throughput comparison for different aggregates specified over 2-hop neighborhoods for \vnma{} overlay 
compared to all pull and all push; we used the write-to-read ratio of 1 over the LiveJournal graph. 
The relative performance of the overlay approach compared to all push or all pull is better for 2-hop aggregates than 1-hop aggregate, which can be attributed
to better sharing opportunities in such queries.

\topicu{Latency}
Figure \ref{fig:real-baseline-latency-scaleup}(c) shows the {\em worst case}, {\em 95th percentile}, and {\em average} latency
for the read queries for \topka{} as the push cost to pull cost ratio is varied. Here we used the network 
packet trace EPA-HTTP to simulate read/write activity. Since the number of distinct IP addresses in the
trace is much smaller than the number of nodes in the (LiveJournal) graph, we randomly split the trace
for each IP address among a set of nodes in the graph.
%
We eliminated contention by ensuring that each query or update runs in isolation.
As we can see, increasing the {\em pull cost} bring down the read latencies,
as {\em pushes} get favored while making dataflow decision. We also note that the worst-case
latencies in our system are quite low.

\topicu{Parallelism} 
Figure \ref{fig:real-baseline-latency-scaleup}(d) shows how the throughput varies as we increase the number of threads
serving the read and write requests for the three approaches; we use the \topka{} query over the
LiveJournal graph, with write-to-read ratio of 1. Because of the synchronization overheads, we do not
see perfect scaleup (note that the $y$-axis is in log-scale); for all three approaches, the throughput increases steadily till about 24
threads, and then plateaus out (our machine has 24 cores with hyperthreading enabled).

\section{Related Work}
\label{sec:rltdwork}
Of the prior work on data stream management, the work on evaluating continuous aggregate
queries over data streams is most closely related to our
work ~\cite{arasu2006cql,pitt-work,streamaggsailesh,streamaggsamrat,streamaggmotwani}. 
Arasu et al. present an SQL-based  continuous query language that supports user defined aggregates and rely on mapping
between streams and relations for efficient implementation~\cite{arasu2006cql}. Krishnamurthy et al.~\cite{streamaggsailesh} and Wang et
al.~\cite{streamaggsamrat} present techniques for sharing work across
different queries with different sliding windows. Al Moakar et al.~\cite{pitt-work} generalize that and 
propose a 3-level aggregation overlay. 
However, the sharing opportunities in ego-centric aggregate 
computation over graphs are fundamentally different and have not been studied in that prior work. 
Further, most of the prior work on evaluating continuous aggregates has only considered the 
{\em all-push} model of query evaluation. 

There has also been much work on aggregate computation in sensor networks and distributed
databases, some of which has considered sharing of partial aggregates (e.g.,~\cite{trigoni,manytomany,tag}).
However the primary 
optimization goal in that work has been minimizing communication cost during distributed execution, 
and hence the techniques developed are quite different.
There is another line of work that deals with aggregation over multiple streams mainly to support monitoring style queries~\cite{distmonhellerstein,distmonolston,distmondivesh}. 
Most of those systems either support distributed monitoring queries by aggregating over multiple streams, or provide
efficient techniques to evaluate multiple aggregate queries on a single
stream~\cite{streamaggsailesh,streamaggsamrat,streamaggsam,streamaggmotwani}. Almost all of the works
discussed so far either focus on static analysis of graphs or aim to do efficient distribution of
data items in order to perform distributed monitoring, hence not really suitable for answering
on-demand neighborhood based aggregation queries for large dynamic graphs. 
Several lines of work have considered the problems in deciding when to push vs pull based on monitoring read/write
frequencies, in the context of replication in distributed data management systems (e.g.,~\cite{wolfson,mondal2012managing}), and
publish-subscribe systems (e.g.,~\cite{feedingfrenzy}). That work has typically focused on minimizing communication cost in
distributed settings rather that the CPU cost of computation. 

Recently, several researchers have looked at the problem of executing subgraph pattern matching queries over streaming graph
data (e.g.,~\cite{sg1}).
Two extensions to SPARQL have also been proposed in recent work for specifying continuous queries over streaming RDF data~\cite{Barbieri:2010:EEC:1739041.1739095,Anicic:2011:EUL:1963405.1963495}. There is also much work on
streaming algorithms for specific problems 
(e.g., {\em counting triangles}~\cite{ct1,ct2}, PageRank computation~\cite{DasSarma:2008:EPG:1376916.1376928}, sketching~\cite{Zhao:2011:GQE:2078331.2078335,Aggarwal:2010:DPM:1920841.1920964}, etc.), and on theoretical models and approximation
algorithms~\cite{theory1,theory2,theory3}. However, we are not aware of any work on supporting continuous queries
specified in a high-level declarative query language. 
Two very recent works, Kineograph~\cite{kineograph} and
GraphInc~\cite{graphinc}, also address continuous analytics over graphs. 
However, none of that prior work considers execution of a large number of ego-centric aggregate queries, further
they do not exploit many of the optimization opportunities (e.g., aggressive sharing, pre-computations, adaptivity, etc.) that
are crucial to handle very high data-rate.

Our approach to overlay construction is closely related to the prior work 
on compressing graphs through identifying structures such as cliques and bicliques~\cite{shinglesravi, buehrer2008scalable, navlakha2008graph}. 
As discussed in Section~\ref{sec:ouralgo}, the most closely related work is by Buehrer et al.~\cite{buehrer2008scalable} who 
propose an algorithm to find bicliques and use them to compress graphs. However, they do not consider
exploiting quasi-bicliques, further, their algorithm cannot handle incremental changes to the graph. Our overlay
construction algorithms can be directly applied to the bipartite graph compression problem, as we show, our algorithms
produce better compression than that prior work.
As a theoretical counterpart, the graph compression work by Feder and Motwani ~\cite{graphcompression} presents {\em clique partitioning} technique 
to compress the base graph in order to achieve better performance for some well studied graph algorithms. 

Network analysis, sometimes called {\em network science}, has been a very active area of research
over the last decade, with much work on network evolution and information diffusion models, community
detection, centrality computation, so on. We refer the reader to well-known surveys and textbooks on that
topic (see, e.g.,~\cite{aggarwal2010managing, newman2003structure, scott2011sage,boccaletti2006complex}).
Increasing availability of temporally annotated network data has led many researchers to focus on designing analytical models that capture how a network evolves, with a
primary focus on social networks and the Web (see, e.g.,~\cite{aggarwal2010managing, leskovec2008microscopic, kumar2010structure, mislove2007measurement}). There is also much work on understanding how communities evolve, identifying key individuals, locating hidden
groups, identifying changes, visualizing the temporal evolution, in dynamic
networks.  Graph data mining is another well researched area where the goal is to find relevant structural 
patterns present in the graph ~\cite{graphmining1,graphmining2,graphmining3,graphmining4}.
Most of that prior work, however, focuses on off-line analysis of static datasets.
%

Ego-centric analysis of information networks has been getting increasing attention in recent years in {\em network science}
community; here the main focus is on structural
analysis of a node's neighborhood~\cite{ego1,ego2}
 as well as on answering specialized
pattern matching queries~\cite{egowaala}. In a recent work, Yan et al.~\cite{nbhdaggrICDE} investigate 
neighborhood aggregation
queries aimed at finding top-$k$ nodes (w.r.t. their aggregate values over their $h$-hop neighborhood) 
in the entire graph. They develop pruning techniques by noting that the aggregate values of two adjacent nodes
are similar. However, they focus on static graphs and they do not consider continuous
evaluation of a large number of ego-centric queries simultaneously.
\section{Conclusions}
\label{sec:conclusions}
In this paper, we presented the design of a continuous query processing system to efficiently
process large numbers of ego-centric aggregation queries over highly dynamic, large-scale graphs. Our
definition of an ego-centric aggregation query is very general, and captures a range of querying and
analytics tasks including personalized trend detection, anomaly or event
detection, and even complex real-time analytics over neighborhoods in the graph. 
We proposed a general framework that supports user-defined aggregate queries and enables efficient evaluation of such queries over highly dynamic graphs; 
we also developed novel scalable algorithms for exploiting sharing opportunities and for making dataflow
decisions based on expected activity patterns. Our system is able to handle graphs containing 320M
nodes and edges on a single machine with 64GB of memory, achieving update and query throughputs over 500k/s. With the
large-memory, many-core machines that are available today, we expect such a centralized approach to be
sufficient in most application domains. 
However, our approach is also naturally parallelizable through use of standard graph partitioning-based
techniques. The {\em readers} can be partitioned in a disjoint fashion over a set of
machines, and for each machine, an overlay can be constructed for the readers assigned to that
machine; the writes for each {\em writer} would be sent to all the machines where they are needed.

\vspace{5pt}
\noindent
\textbf{Acknowledgments:}
This work was supported by NSF under grant IIS-1319432, by Air Force Research Lab (AFRL) under
contract FA8750-10-C-0191 and by an IBM Faculty Award.

\vspace{0pt}


%

\end{document}